\documentclass[aps,prb,twocolumn]{revtex4}
\usepackage{graphicx}
\usepackage{dcolumn}
\usepackage{amsmath}
\usepackage{amssymb}
\usepackage{subfigure,amsmath,verbatim,moreverb,bm}
\def\be{\begin{equation}}
\def\ee{\end{equation}}
\def\ber{\begin{eqnarray}}
\def\eer{\end{eqnarray}}
\def\xiv{{\boldmath{\xi}}}

\def\nablabold{\mbox{\boldmath $\nabla$}}
\def\rv{{\bf r}}
\def\gv{{\bf g}}
\def\jv{{\bf j}}
\def\kv{{\bf k}}
\def\qv{{\bf q}}
\def\Av{{\bf A}}
\def\Bv{{\bf B}}
\def\Ev{{\bf E}}

\def\uv{{\bf u}}
\def\vv{{\bf v}}

\def\nn{\nonumber}
\def\xiv{\bm{\xi}}

\begin{document}
\title{Continuum Mechanics for Quantum Many-Body Systems:\\ The Linear Response Regime}
\author {Xianlong Gao}
\affiliation{Department of Physics, Zhejiang Normal University, Jinhua, Zhejiang Province, 321004, China}\author{Jianmin Tao}
\affiliation{Theoretical Division and Center for Nonlinear Studies,
Los Alamos National Laboratory, Los Alamos, New Mexico 87545}
\author{G. Vignale}
\affiliation{Department of Physics, University of Missouri-Columbia,
Columbia, Missouri 65211}
\author{I. V. Tokatly}
\affiliation{IKERBASQUE, Basque Foundation for Science, E-48011, Bilbao, Spain}
\affiliation{ETSF Scientific Development Centre, 
Dpto. F\'isica de Materiales, Universidad del Pa\'is Vasco, Centro de
F\'isica de Materiales CSIC-UPV/EHU-MPC, Av. Tolosa 72, E-20018 San 
Sebasti\'an, Spain}
\affiliation{Moscow Institute of Electronic Technology, Zelenograd, 124498 Russia}
\date{\today}
\begin{abstract}
We derive a closed equation of motion for the current density of an inhomogeneous quantum many-body system under the assumption that the time-dependent wave function can be described  as a geometric deformation of the ground-state wave function.   By describing the many-body system in terms of a single collective field we provide an alternative to traditional approaches, which emphasize one-particle orbitals.  We refer to our approach as continuum mechanics  for quantum many-body systems.  In the linear response regime, the equation of motion for the displacement field becomes a linear  fourth-order integro-differential equation, whose only inputs are the one-particle density matrix and the pair correlation function of the ground-state.  The complexity of this equation remains essentially unchanged as the number of particles increases.  We show that our equation of motion is a hermitian eigenvalue problem, which admits a complete set of orthonormal eigenfunctions under a scalar product that involves the ground-state density.  Further, we show that the excitation energies derived from this approach satisfy a sum rule which guarantees the exactness of the integrated spectral strength.  Our formulation becomes exact for systems consisting of a single particle, and for any many-body system in the high-frequency limit.  The theory is illustrated by explicit calculations for simple one- and two-particle systems.\\
\hspace*{7.0cm}
\par
\end{abstract}
\maketitle
\section{Introduction}
The dynamics of quantum many-particle systems poses a major challenge to computational physicists and chemists.  In the study of ground-state properties one can rely on a variational principle, which enables a variety of powerful statistical methods (in addition to exact diagonalization)  such as the quantum variational Monte Carlo method and the diffusion Monte Carlo method.\cite{MonteCarlo}
In time-dependent situations, the absence of a practical variational principle has greatly hindered the development of equally powerful methods.    Yet it is hard to overestimate the importance of developing effective techniques to tackle the quantum dynamical problem.  Such a technique could allow, for example, to follow in real time the evolution of chemical reactions, ionization and collision processes.

 One of the most successful computational methods developed to date  is the time-dependent density functional theory (TDDFT), or its more recent version -- time-dependent current density functional theory (TDCDFT). \cite{grossbook}    In this approach, the interacting electronic system is treated as a noninteracting electronic system subjected to an effective scalar potential (a vector potential in TDCDFT) which is self-consistently determined by the electronic density (or by the current density).\cite{Gross96,Casida95}  Thus, one avoids the formidable problem of solving the time-dependent Schr\"odinger equation for the many-body wave function.  Even this simplified problem, however, is quite complex, since it involves the determination of $N$ time-dependent single particle orbitals -- one for each particle.   Furthermore, there are features such as multi-particle excitations \cite{Maitra04} and dispersion forces\cite{Dobson98}  that are very difficult to treat within the conventional approximation schemes.

An alternative approach, which actually dates back to the early days of the quantum theory, attempts to calculate the collective variables of interest, density and current, without appealing to the underlying wave function.\cite{TerHaar,Madelung1927,Bloch1933}  This approach we call ``quantum continuum mechanics" (QCM), because in analogy with classical theories of continuous media (elasticity and hydrodynamics),  it attempts to describe the quantum many-body system without explicit reference to the individual particles of which the system is constituted.\cite{GhoDeb1982}

That such a description is possible is guaranteed by the very same theorems that lie at the foundation of TDDFT and TDCDFT.\cite{rg,vanLeeuwen99}  Indeed, consider a system of particles of mass $m$ described by the time-dependent hamiltonian
\be\label{defH}
\hat H(t) = \hat H_0+\int d\rv \hat n(\rv)V_1(\rv,t)
\ee
where  
\be\label{defH0}
\hat H_0 = \hat T +\hat W+\hat V_0
\ee
is the sum of kinetic energy ($\hat T$),  interaction potential energy ($\hat W$),  and the energy associated with an external {\it static} potential $V_0(\rv)$,
\be\label{defV0}
\hat V_0 = \int d\rv V_0(\rv)\hat n(\rv)\,,
\ee
where $\hat n(\rv)$ is the particle density operator.  $V_1(\rv,t)$ is an external time-dependent potential. 

The exact Heisenberg equations of motion for the density and the current density operators, averaged over the quantum state, lead to equations of motion for the average particle density $n(\rv,t)$ and the average particle current density $\jv(\rv,t)$:
\be  \label{Continuity}
\partial_t n(\rv,t) = -\partial_\mu j_\mu (\rv,t)
\ee
and
\ber\label{ForceBalance} 
m\partial_t j_\mu(\rv,t)=
&-&n(\rv,t) \partial_\mu [V_0(\rv)+V_1(\rv,t)]\nn\\
&-&\partial_\nu P_{\mu\nu}(\rv,t)\,, 
\eer
where $\partial_t$  denotes the partial derivative with respect to time and $\partial_\nu$ is a short-hand for the derivative with respect to the cartesian component $\nu$ of the position vector $\rv$.  Here and in the following we adopt the convention that repeated indices are summed over.    These equations simply express the local conservation of particle number (Eq.~(\ref{Continuity})) and momentum (Eq.~(\ref{ForceBalance})).  The key quantity on the right hand side of Eq.~(\ref{ForceBalance}) is the {\it stress tensor} $P_{\mu\nu}(\rv,t)$ --  a symmetric tensor which will be defined in the next section as the expectation value of a hermitian operator,  and whose divergence with respect to one of the indices yields the {\it force density} arising from internal quantum-kinetic and interaction effects. 

Now the Runge-Gross theorem of TDDFT guarantees that the stress tensor, like every other observable of the system, is a functional of the current density  and of the initial
quantum state.  Thus,  Eq.~(\ref{ForceBalance}) is in principle a
closed equation of motion for $\jv$ -- the only missing piece of information 
being the {\it explicit} expression for $P_{\mu\nu}$ in terms of the
current density.

In recent years much effort has been devoted to constructing an approximate QCM\cite{Zaremba94,ConVig1999,TokPanPRB1999,TokPanPRB2000,DobLe,tokatly2005a,tokatly2005b,tokatly2007,tvt07} and several applications have appeared in the literature (see Ref.~\onlinecite{Applications} for some representative examples).   All approximation schemes so far have been based on the local density approximation and generalizations thereof.   The objective of this paper is to derive and discuss an approximate expression for
$P_{\mu\nu}(\rv,t)$, and, more importantly, for the associated force density $-\partial_\nu P_{\mu\nu}(\rv,t)$,  as functionals of the current density and the initial state.
We will do this in the {\it linear response regime}, i.e. for systems that start from the ground-state of the static hamiltonian $\hat H_0$ and perform small-amplitude oscillations about it.   The external potential $V_1(\rv,t)$ will be treated as a small perturbation.  In this regime the equations of motion~(\ref{Continuity}) and (\ref{ForceBalance})  are conveniently expressed in terms of the displacement field $\uv(\rv,t)$, defined by the relation
\be
 \jv(\rv,t) = n_0(\rv) \partial_t \uv(\rv,t)\,,
\ee
where $n_0(\rv)$ is the ground-state density.  It is also convenient to write the density and the stress tensor as sums of a large ground-state component and a small time-dependent part, in the following manner
\ber
n(\rv,t)&=&n_0(\rv)+n_1(\rv,t)\,,\nn\\
P_{\mu\nu}(\rv,t)&=&P_{\mu\nu,0}(\rv)+P_{\mu\nu,1}(\rv,t)\,,
\eer
where the equilibrium components, marked by the subscript $0$, satisfy the equilibrium condition
\be\label{equilibrium1}
n_0(\rv)\partial_\mu V_0(\rv)+\partial_\nu P_{\mu\nu,0}(\rv)=0\,.
\ee
Then the two equations ~(\ref{Continuity}) and (\ref{ForceBalance})  take the form
\be
n_1(\rv,t) =-\partial_\mu [n_0(\rv) u_\mu(\rv,t)]\,,
\ee
and
\ber\label{eom.1}
&&m n_0(\rv)\partial_t^2 u_\mu (\rv,t)=-n_0(\rv)\partial_\mu V_1(\rv,t)  \nn\\&&-n_1(\rv,t)\partial_\mu V_0(\rv)-\partial_\nu P_{\mu\nu,1}(\rv,t)\,.
\eer
Our  task is to find an expression for the force density $\partial_\nu P_{\mu\nu,1}(\rv,t)$ as a linear functional of $\uv(\rv,t)$.  If this can be achieved, then the excitation energies of the system will be obtained from the frequencies of the time-periodic solutions of Eq.~(\ref{eom.1}) in the absence of external field (i.e., with $V_1=0$).   

It is easy to see that the spatial dependence of these solutions will be proportional to the matrix element of the current density operator between the ground-state and the excited state in question.  This is because, in a many-body system with stationary states $|\psi_0\rangle, |\psi_1\rangle, ..., |\psi_n\rangle, ...$  ($|\psi_0\rangle$ is the ground-state),  and corresponding energies $E_0,E_1, ..., E_n...$,  the n-th linear excitation is described by the time-dependent state
\be
|\psi_0\rangle e^{-iE_0t}+\varepsilon|\psi_n\rangle e^{-iE_nt}\,,
\ee
where $\varepsilon$ is an arbitrarily small ``mixing parameter".  The expectation value of the current density operator in this state is
\be
\jv(\rv,t) = \varepsilon\langle \psi_0|\hat \jv(\rv)|\psi_n\rangle e^{-i(E_n-E_0)t}+ c.c\,.
\ee
Thus, in principle, almost all the excitation energies $(E_n-E_0)$ of the system can be obtained by Fourier-analyzing the displacement field -- the only exception being those excitations that  are not connected to the ground-state by a finite matrix element of the current-density operator.

In this paper we will introduce an approximate expression for the force density 
\be\label{ForceDensity1}
{F}_{\mu,1}(\rv,t) \equiv -n_1(\rv,t)\partial_\mu V_0(\rv)-\partial_\nu P_{\mu\nu,1}(\rv,t)\,,
\ee 
which appears on the right hand side of Eq.~(\ref{eom.1}), as a linear functional of $\uv(\rv,t)$.  The expression  will be presented in terms of the functional 
\be\label{defEu}
E[\uv] \equiv \langle \psi_0[\uv]|\hat H_0|\psi_0[\uv]\rangle\,,
\ee
which is the energy of the distorted ground-state $|\psi_0[\uv]\rangle$,  obtained from the undistorted ground-state $|\psi_0\rangle$ by virtually displacing the volume element  located at $\rv$ to a new position $\rv+\uv(\rv,t)$.  More precisely, we will show that the equation of motion for $\uv$ takes the form:
\ber\label{eom.simple}
m n_0(\rv)\partial_t^2 \uv (\rv,t)  = -n_0(\rv)\nablabold V_1(\rv,t)-\frac{\delta E_2[\uv]}{\delta \uv (\rv,t)}\,,\nn\\
\eer
where $E_2[\uv]$ is the second order term in the expansion of $E[\uv]$ in powers of $\uv$.
The functional $E_2[\uv]$ has an exact expression in terms of the  one-particle density matrix and the pair correlation function of the {\it ground-state}, which is  a major simplification, since ground-state properties, unlike time-dependent properties,  are accessible to computation by a variety of numerical and analytical methods.   

%
Furthermore, we will show that the kinetic part of the force density functional $\delta E_2[\uv]/\delta \uv$  is local, in the sense that it depends only on a finite number of spatial derivatives (up to the fourth) of the displacement field at a given position.    Thus, our equation of motion reduces to a fourth-order differential equation for $\uv$ when interaction effects are neglected.  The inclusion of interaction effects  leads to the appearance of nonlocal contributions to the energy, and the equation of motion becomes  a fourth-order {\it integro-differential} equation for the displacement field.  However, the complexity of this equation remains essentially unchanged as the number of particles increases.

Our equation of motion has two especially appealing features:   (i) it is exact for  one-electron systems at all frequencies and (ii) it can be physically justified for generic many-electron systems at high frequency or, more generally, at all frequencies for which a collective description of the motion is plausible.  Thus the range of frequencies for which our approximation makes sense is expected to be wider in strongly correlated systems than in weakly correlated ones. 
  
We discuss several qualitative features of our equation (uniform electron gas limit, harmonic potential theorem) and present its solution in simple one- and two-electron models, where the results can be checked against exact calculations.  The results are encouraging.  Although we are not able to resolve all the different excitation energies of the models under study, we find that groups of excitation characterized by similar displacement fields are represented by a single mode of an average frequency, in such a way that the spectral strength of this mode equals the sum of the spectral strengths of all the excitations in the group.  In this sense our approximation can be viewed as a (considerable) refinement and extension of the traditional single-mode approximation for the homogeneous electron gas to strongly inhomogeneous quantum systems.  In spite of the somewhat limited range of validity of the present treatment (the linear response regime), we feel that this is an important first step in a direction that might eventually lead to the construction of useful force density functionals for far-from-equilibrium processes.

This paper is organized as follows.  In Section II we present a complete derivation of the linearized equation of motion for the displacement field.  We begin by deriving a formally exact expression for the force density (Section II A), on which we perform the  ``elastic approximation"  (Section II B).    The expression for the force density in the elastic approximation is worked out in sections II C (kinetic part) and II D (potential part).  A simplified form of the equation of motion, valid for one-dimensional systems, is presented in Section II E.  Appendixes A through C provide supporting material for this part.  In Section III we discuss the relation between quantum continuum mechanics and time-dependent current density functional theory.  In section IV we show how the linear equation of motion derived in Section II leads to an eigenvalue problem for the excitation energies.  In Section IV A we demonstrate the hermiticity of this eigenvalue problem and the positive-definiteness of the eigenvalues.  In Section IV B we connect the eigenvalue problem to the high-frequency limit of the linear response theory.  In Section IV C we prove that the first moment of the current excitation spectrum obtained from the solution of our eigenvalue problem is exact.  Appendixes D and E  contain supporting material for this part.
In Section V we present a few simple applications of our theory for the excitations of (i) a homogeneous electron gas (Section V A) (ii) the linear harmonic oscillator and the  hydrogen atom  (Section V B), and  (iii) a system of two-electrons in a one-dimensional parabolic potential interacting via a soft Coulomb potential.  The analytic solution of the last model in the strong correlation regime is featured in Appendix F.   Finally, Section VI contains our summary and a few speculations about future applications of the theory.

\section{Linearized equation of motion}
\subsection{Derivation of the force density}
In this section we undertake the construction of an  approximate expression for the force density, Eq.~(\ref{ForceDensity1}),  as a linear functional of $\uv$.    The stress tensor $P_{\mu\nu}(\rv,t)$, whose divergence determines the force density,  is defined as the expectation value of the stress tensor {\it operator}  $\hat P_{\mu\nu}(\rv)$ in the evolving quantum state $|\psi(t)\rangle$:
\be\label{DefP}
P_{\mu\nu}(\rv,t)= \langle \psi (t)| \hat P_{\mu\nu}(\rv)|\psi(t)\rangle \,.
\ee
An exact and unambiguous expression for the operator $\hat P_{\mu\nu}(\rv)$ in an arbitrary system of coordinates is obtained by considering the  universal  many-body Hamiltonian 
\be \label{H-Universal}
\hat H_u =\hat T +\hat W
\ee 
(external potential {\it not} included) in the presence of a ``metric tensor field"  $g_{\mu\nu}(\rv)$.  As is well known~\cite{LL2},  the metric tensor $g_{\mu\nu}(\rv)$  allows us to express the length $ds$ of an infinitesimal displacement  from $\rv$ to  $\rv+d\rv$ in terms the corresponding  increments  of the coordinates $dr_\mu$:
\be\label{def-metrics}
ds^2 = g_{\mu\nu}(\rv) dr_{\mu}dr_{\nu}\,.
\ee
In ordinary Euclidean space and in cartesian coordinates $g_{\mu\nu}(\rv)=\delta_{\mu\nu}$, independent of position.   In general, however, a non-Euclidean space is characterized by a position-dependent, symmetric  $g_{\mu\nu}(\rv)$.  A non-Euclidean metric can also be generated by a change of coordinates in an Euclidean space, as we will see shortly.  
As an important technical point we also introduce the tensor $g^{\mu\nu}$  as the {\it inverse} of $g_{\mu\nu}$, and we define $g$ as the {\it determinant} of $g_{\mu\nu}$ (so $g^{-1}$ is the determinant of $g^{\mu\nu}$).  

The hamiltonian $\hat H_u$ undergoes the following changes in the presence of a non-trivial metrics.  First, the laplacian operator $\partial_\mu\partial_\mu$ for the kinetic energy is replaced by
\be\label{Laplacian}
 \frac{1}{\sqrt{g}} \partial_\mu \sqrt{g} g^{\mu\nu}\partial_\nu\,. 
 \ee
Second, the Euclidean distance between two points (which controls the interaction energy) is replaced by the non-Euclidean length of the shortest path (geodesic) connecting the two points.  We denote by $\hat H_u[{\bf g}]$ the Hamiltonian in the presence of the metric field $g_{\mu\nu}$.    Then the stress tensor operator is defined as the first variation of $\hat H_u[{\bf g}]$  under an instantaneous virtual variation of the metric tensor $g_{\mu\nu}(\rv)$, i.e.,
\be\label{defP}
\hat P_{\mu\nu}(\rv)\equiv\frac{2}{\sqrt{g}}\frac{\delta \hat H_u[{\bf g}]}{\delta g^{\mu\nu}(\rv)} \,.
\ee
The first-order change in the Hamiltonian due to a change $\delta g_{\mu\nu}$ in the metric tensor is given by
\be
\hat H_u[{\bf g}]\to \hat H_u[{\bf g}]+ \int d\rv \frac{\sqrt{g(\rv)}}{2} \hat P_{\mu\nu}(\rv)\delta g^{\mu\nu}(\rv)\,.
\ee
Notice that the stress tensor operator defined in this manner is itself a functional of the metrics.  This definition is completely analogous to the standard definition of the current density operator as the derivative of the Hamiltonian with respect to a vector potential.  An explicit expression for $\hat P_{\mu\nu}$ in Euclidean metrics is reported for completeness in Appendix A (see also Refs~\onlinecite{tokatly2005a}, \onlinecite{tokatly2005b}, \onlinecite{Puff-Gillis} and \onlinecite{Nielsen-Martin}). We note that the definition of the quantum mechanical stress tensor via the variational derivative with respect to the metric tensor has been also employed in Ref.~\onlinecite{Rappe}.

We will now focus on the calculation of $P_{\mu\nu,1}$ -- the correction to $P_{\mu\nu}$ of first order in $\uv$.  In order to express $P_{\mu\nu,1}(\rv,t)$  and its divergence as functionals of the displacement field we resort to Tokatly's recent formulation of quantum dynamics in the co-moving reference frame.\cite{tokatly2005a,tokatly2005b,tokatly2007,tvt07}  The co-moving frame is an accelerated reference frame which, at each point and each time, moves with the velocity of the volume element of the fluid at that point and that time,  so that the density is constant and equal to the ground-state density, while the current density is zero.  The time-dependent transformation from the laboratory frame (coordinates $\rv$) to the  co-moving frame (coordinates $\xiv$) is defined by the solution of the equation
\be\label{LagrangianEOM}
\partial_t \rv(t) =  \vv(\rv,t)\,,~~~~~~~~\rv(0) = \xiv\,,
\ee
where $\vv(\rv,t)= \frac{\jv(\rv,t)}{n(\rv,t)}$ is the velocity field.  In the linear response regime the velocity is approximated as $\jv(\rv,t)/n_0(\rv)$, where $n_0(\rv)$ is the ground-state density. In this regime we can write
\be \label{CoordinateTransformation}
\rv(t) = \xiv+\uv(\xiv,t)
\ee
where $\uv(\xiv,t)$ is the (small) displacement of a fluid element for its initial position $\xiv$.    Expressing $ds^2 = d\rv \cdot d\rv$ in terms of the new coordinates $\xiv$ and making use of Eq.~(\ref{def-metrics})  we see that the metric tensor in the co-moving frame is given by
\be\label{MetricTensor}
g_{\mu\nu}(\xiv,t)  = \frac{\partial r_{\alpha}}{\partial \xi_{\mu}}\frac{\partial r_{\alpha}}{\partial \xi_{\nu}}\,.
\ee
From Eq.~(\ref{CoordinateTransformation}), to first order in the displacement field, we immediately get
\begin{eqnarray}\label{metric1}
g_{\mu\nu} = \delta_{\mu\nu}+2u_{\mu\nu}~,
\end{eqnarray}
and 
\begin{eqnarray}\label{metric2}
g^{\mu\nu} = \delta_{\mu\nu}-2u_{\mu\nu}~,
\end{eqnarray}
where
\be\label{StrainTensor}
u_{\mu \nu} \equiv \frac{1}{2}\left(\partial_\nu u_{\mu} + \partial_\mu u_{\nu} \right)
\ee
is the {\it strain tensor}.
Also to first order in $\uv$ the determinant of the metric tensor is easily seen to be
\be\label{g-determinant}
g \simeq 1+2 \nablabold\cdot\uv\,,
\ee
so that, for example,  $g^{-1/2}\simeq 1-\nablabold \cdot \uv$. 
In view of these relations we will, in the rest of this paper, replace $\hat H_u[\gv]$ by $\hat H_u[\uv]$, with the understanding that $\uv$ completely determines the metrics.   We also notice that, by virtue of Eq.~(\ref{metric2}), we have
\be
\frac{\delta \hat H_u[\gv]}{\delta g^{\mu\nu}(\rv)} =  -\frac{1}{2}\frac{\delta \hat H_u[\uv]}{\delta u_{\mu\nu}(\rv)}\,.
\ee

The main reason for introducing the co-moving reference frame is that in this frame we can make a simple approximation, which enormously simplifies the task of linearizing the stress tensor.  This will be discussed in the next section.  For the time being we proceed in a formally exact manner.  To begin with, we observe that  the general relation between the stress tensor in the lab frame and that in the co-moving frame is
\be\label{TensorTransformation}
P_{\mu\nu}(\rv,t)= (\partial_\mu \xi_\alpha) (\partial_\nu \xi_\beta) \tilde P_{\alpha\beta}(\xiv(\rv,t),t)\,,
\ee
where
\be\label{DefPtilde}
\tilde P_{\mu\nu}(\rv,t)= -\frac{1}{\sqrt{g(\rv,t)}}\langle \tilde \psi (t)| \frac{\delta \hat H_u[\uv]}{\delta u_{\mu\nu}(\rv)}|\tilde \psi(t)\rangle \,,
\ee
where $|\tilde \psi (t)\rangle$ is the quantum state in the co-moving frame.\footnote{To be completely accurate, we point out that the hamiltonian $\hat{\tilde H}(t)$, which governs the time evolution of the quantum state $|\tilde \psi(t)\rangle$ in the co-moving reference frame, does not coincide with the instantaneously deformed hamiltonian $\hat H[\uv]$.  The difference arises from the fact that the coordinate transformation to the co-moving frame is time-dependent, and this generates an additional vector potential (also a functional of $\uv$), which guarantees the vanishing of the current density in the co-moving frame.  Happily, this vector potential becomes irrelevant in the high-frequency limit, and therefore does not contribute to the elastic approximation proposed in this paper.} 

After expanding the stress tensor in the co-moving frame to first order in the displacement field,
\be
\tilde P_{\mu\nu}(\xiv,t) = P_{\mu\nu,0}(\xiv)+\tilde P_{\mu\nu,1}(\xiv,t)\,,
\ee
it is easy to see that in the lab frame we have
\be\label{tensor1.1}
P_{\mu\nu,1}=\tilde P_{\mu\nu,1} -\uv\cdot\nablabold P_{\mu\nu,0}
- (\partial_\mu u_\alpha) P_{\alpha\nu,0}-(\partial_\nu u_\alpha) P_{\mu\alpha,0}\,.
\ee
From this we get
\ber\label{fv1.1}
\partial_\nu P_{\mu\nu,1}&=&\partial_\nu [\tilde P_{\mu\nu,1}-(\partial_\mu u_\alpha) P_{\alpha\nu,0}-(\partial_\nu u_\alpha) P_{\mu\alpha,0}]\nn\\
&-&\partial_\nu (\uv\cdot  \nablabold P_{\mu\nu,0})\,,
\eer
and, after some algebra, 
\ber\label{fv1.2}
&&n_1\partial_\mu V_0+ \partial_\nu P_{\mu\nu,1} = n_0 \uv \cdot \nablabold \partial_\mu V_0\nn\\
&+&\partial_\nu [\tilde P_{\mu\nu,1}+(\nablabold\cdot\uv)P_{\mu\nu,0}-(\partial_\mu u_\alpha) P_{\alpha\nu,0}-2 u_{\alpha\nu}  P_{\mu\alpha,0}]\nn\\
\eer
where we have made use of the equilibrium condition~(\ref{equilibrium1}) and the definition ~(\ref{StrainTensor}) of the strain tensor.

It is convenient at this point to introduce the first-order stress force density 
\be
\label{F1}
{\cal F}_{\mu,1}\equiv -\partial_\nu [\tilde P_{\mu\nu,1}+(\nablabold\cdot\uv)P_{\mu\nu,0}-(\partial_\mu u_\alpha) P_{\alpha\nu,0}-2 u_{\alpha\nu}  P_{\mu\alpha,0}]\,
\ee
so that the equation of motion~(\ref{eom.1})  takes the form
\ber\label{eom.3}
m\partial_t^2u_\mu (\rv,t) + \uv\cdot\nablabold  \partial_\mu V_0 = 
\frac{{\cal F}_{\mu,1}(\rv,t)}{n_0(\rv)} -\partial_\mu V_1(\rv,t)\,.\nn\\
\eer 

Finally, it is possible to prove (see Appendix B) that the first order force density is exactly given by the expression 
\ber\label{IlyasIdentity}
{\cal F}_{\mu,1}(\rv,t)= -\langle \tilde \psi(t)| \frac{\delta \hat H_u[\uv]}{\delta u_\mu(\rv)}|\tilde\psi(t)\rangle\Big\vert_1\,,
\eer
where $\frac{\delta\hat H_u[\uv]}{\delta u_\mu(\rv)}$ is its functional derivative calculated with respect to a {\it virtual} (time-independent) variation of the displacement field.  The vertical bar $\Big\vert_1$ mandates that we keep only the first-order in $\uv$ part of the bracketed expression. 

With the help of this identity we see that the equation of motion for the displacement field takes the form
\ber\label{eom.exact}
&&m\partial_t^2u_\mu (\rv,t) + \uv\cdot\nablabold  \partial_\mu V_0=\nn\\
&-&\frac{1}{n_0(\rv)}\langle \tilde \psi(t)| \frac{\delta\hat H_u[\uv]}{\delta u_\mu (\rv)}|\tilde\psi(t)\rangle\Big\vert_1 -\partial_\mu V_1(\rv,t)\,.\nn\\
\eer 
The same result could have been derived almost immediately by using the more sophisticated machinery of the generally covariant Lagrangian formalism introduced in Ref.~\onlinecite{tokatly2007}. In fact, Eq.~(\ref{eom.exact}) is simply a linerized version of the equation of motion for an infinitesimal fluid element, Eq.~(39) of Ref.~\onlinecite{tokatly2007}.

As a reality check, let us ask ourselves whether the system can support excitations in which the displacement field is uniform in space: $\uv(\rv,t)=\uv(t)$.  Clearly in this case the strain vanishes and there is no change in metrics, so  $\delta \hat H_u[\uv]/\delta \uv$ is null.  As a result, after setting the external field $V_1 =0$ we get the equation
\be
 m\partial_t^2 u_\mu (t)+\uv(t) \cdot\nablabold  \partial_\mu V_0 =0
 \ee
which has the solution $u_\mu(t) \propto \cos (\omega t+\phi)$ if and only if the potential is of the harmonic form $V_0(\rv)=\frac{1}{2} m\omega^2 r^2$.  This is just a statement of the {\it harmonic potential theorem}~\cite{Dobson} according to which a many-body system  in a harmonic potential performs a rigid simple harmonic motion with frequency $\omega$ imposed by the curvature of the harmonic potential. We have now shown that the harmonic potential is the only potential with this property.

\subsection{The elastic approximation}

Equation~(\ref{eom.exact}) is formally exact, but it still contains the  time-dependent state $|\tilde \psi(t)\rangle$, which of course is not known.    In spite of the simple behavior of the density (constant) and the current density (null), the evolution of the many-body wave function in the co-moving frame is far from trivial.  Nevertheless, a simple and physically appealing approximation suggests itself.  Namely, we assume that the wave function $\tilde \psi$ is time-independent (just as the density), and coincides with the ground-state wave function of the laboratory frame ($\psi_0$)  evaluated at the coordinates $\xiv$ of the co-moving frame:
\be\label{ElasticApproximation}
\tilde \psi(\xiv_1,...,\xiv_N,t)  \simeq \psi_0(\xiv_1,...,\xiv_N)\,.
\ee 

 The physical idea behind this approximation is that the time evolution of the wave function in the laboratory frame can be approximated as a continuously evolving elastic deformation of the ground-state wave function.  Such a deformation affects all the particles simultaneously and instantaneously.  The burden of describing the time-evolution of the system is entirely placed on the time-dependent geometry (i.e. the time-dependent relation between $\xiv$ and $\rv$),  while the wave function itself remains independent of time.   
 
 What is {\it lost} in this approximation is the fact that in the actual time evolution the system will undergo internal relaxation in order to optimize the correlations between the particles.  In other words, the probability of finding the particles in a certain configuration $\rv_1,...\rv_N$ at time $t$ is {\it not} strictly determined by the probability that those particles were initially  in the configuration $\xiv_1,...,\xiv_N$ from which $\rv_1,...\rv_N$ evolve according to Eqs.~(\ref{LagrangianEOM}) and (\ref{CoordinateTransformation}).  
 However, our approximation should always be valid at sufficiently high frequency, i.e., when the evolution of the geometry is very fast on the scale of the characteristic response times of the system.  
 
 The equation of motion resulting from  the elastic approximation is also strictly valid (and therefore, not an approximation at all) for any one-particle system, because in this case the wave function {\it is} completely determined by the displacement field and there is no room for internal relaxation.  Finally, our equation of motion is also strictly valid for non-interacting Bose systems  in the ground-state (since these systems behave like a single particle), and for non-interacting Fermi systems consisting of at most two particles of opposite spins in the same orbital (since these behave like non-interacting Bosons).  In all other cases -- including the apparently simple case of a non-interacting many-fermion system  -- the appropriateness of the elastic approximation must be assessed {\it a posteriori} and may depend on the objective  of the calculation.  In general, we can only say that the elastic approximation is expected to work better for collective (many-particle)  excitations than for single particle excitations, and better for strongly correlated many-body systems (which exhibit bosonic behavior) than for weakly correlated systems.

It is important to appreciate the profound difference that exists between the present approximation and another common approximation which also entails an instantaneous response to a time-dependent field: the {\it adiabatic approximation}.  In the adiabatic approximation one assumes that the system  remains in the instantaneous ground-state of the hamiltonian $\hat H(t)$ -- an assumption that is justified only if the time evolution is {\it slow} on the scale of the characteristic response time of the system.  This is exactly the opposite of the regime of validity of the present approximation.  The geometrically distorted wave function $\tilde \psi$ is not at all close to the instantaneous ground-state of $\hat H(t)$.  Rather, it is the ground-state of the ``deformed hamiltonian" $\hat H_0[\uv]$ which is obtained from the initial-time Hamiltonian $\hat H_0$ by a coordinate transformation -- indeed an elastic deformation.

As anticipated in the foregoing discussion the elastic approximation paves the way for a relatively simple calculation of the complicated expression that appears in the second line of Eq.~(\ref{eom.exact}).  Namely, thanks to the fact that $\tilde \psi = \psi_0$ is independent of the displacement field we can take the functional derivative of Eq.~(\ref{IlyasIdentity}) {\it after} taking the average and we arrive at 

\ber\label{IlyasIdentity2}
{\cal F}_{\mu,1}(\rv,t)=-\left.\frac{\delta E_u[\uv]}{\delta u_\mu (\rv)}\right\vert_1\,,
\eer
where 
\be
E_u[\uv]\equiv \langle \psi_0|\hat H_u[\uv]|\psi_0\rangle\,.
\ee
We further observe that
\be
 \uv\cdot\nablabold  \partial_\mu V_0(\rv)=\frac{1}{n_0(\rv)} \frac{\delta V_0[\uv]}{\delta u_\mu(\rv)}\Big\vert_1\,,
 \ee
where 
\be
V_0[\uv]\equiv\int d\rv V_0(\rv+\uv(\rv)) n_0(\rv)\,.
 \ee
 is the external potential energy of the distorted ground-state.


 Putting all together  we arrive at the elegant result:
\ber\label{eom.exact2}
m\partial_t^2u_\mu (\rv,t) =-\frac{1}{n_0(\rv)}\frac{\delta E_2[\uv]}{\delta u_\mu(\rv)}-\partial_\mu V_1(\rv,t)\,.\nn\\
\eer 
where $E_2[\uv]$ is the second-order term in the expansion of the {\it total energy} 
\be
E[\uv]\equiv E_u[\uv]+V_{0}[\uv]
\ee
 of the distorted ground state.
Equivalently, $E[\uv]$ can be obtained  as the expectation value of the original hamiltonian $\hat H_0$ in the distorted ground state
\ber
\psi_0[\uv](\rv_1,...,\rv_N)&=&\psi_0(\xiv_1,...,\xiv_N) \prod_{i=1}^{N}g^{-1/4}(\xiv_i)\,,
\eer
where $\xiv_i=\rv_i-\uv(\rv_i)$  and the last factor on the right hand side is for normalization. 
This proves that $E[\uv]-E[{\bf 0}]$ is a positive definite quantity since $E[{\bf 0}]$ is the ground-state energy of $\hat H_0$ while $E[\uv]$ is the expectation value of $\hat H_0$ in a state that is not the ground-state. 

\subsection{Calculation of  $\delta E_2[\uv]/\delta\uv$ -- kinetic part}

The evaluation of the distorted ground-state energy $E_2[\uv]$  is in principle straightforward if the exact one-particle and two-particle density matrices of the ground-state are known.  In this section we focus on the construction of the kinetic contribution, which, as we will show,  leads to a {\it local} equation of motion,  which involves only a finite number of derivatives (up to the fourth) of the displacement field.
For a calculation of the kinetic contribution to the elastic energy only the one-particle density matrix
\be\label{defrho}
\rho(\rv,\rv')\equiv\langle \psi_0|\hat \Psi^\dagger(\rv)\hat \Psi(\rv')|\psi_0\rangle
\ee
is needed. 
\begin{widetext}
The kinetic energy  of the distorted state is
\be
\label{Tfull}
T[\uv] = \frac{1}{2m}\int d\rv  \sqrt{g} g^{\mu\nu}\partial_\mu\partial_\nu^\prime [g^{-1/4}(\rv) g^{-1/4}(\rv') \rho(\rv,\rv')]_{\rv=\rv'}\,,
\ee
which reduces to the kinetic energy of the ground-state when $\uv=0$.  Expanding the above expression to second order in $\uv$ we arrive, after some laborious algebra (see Appendix C for the derivation) to the following expression
\ber\label{T2uv}
T_2[\uv] &=& \int d\rv \left \{ 2T_{\mu\nu,0}\left[u_{\mu\alpha}u_{\nu\alpha}-\frac{1}{4}(\partial_\mu u_\alpha)(\partial_\nu u_\alpha)\right]
+\frac{n_0}{8m} (\partial_\mu u_{\nu\nu})(\partial_\mu u_{\nu\nu})\right.\nn\\&+&\left.\frac{n_0}{2m}[(\partial_\mu u_{\nu\alpha})(\partial_\mu u_{\nu\alpha})-
(\partial_\mu u_{\nu\mu})(\partial_\nu u_{\alpha\alpha})]\right\}\,,
\eer
where
\be\label{Tmunu0}
T_{\mu\nu,0}=\frac{1}{2m}\left(\partial_\mu\partial_\nu^\prime+\partial_\nu\partial_\mu^\prime\right) \rho(\rv,\rv')\Big\vert_{\rv=\rv'} - \frac{1}{4m}\nabla^2 n_0\delta_{\mu\nu}
\ee
is the equilibrium stress tensor.
Notice that $T_2[\uv]$ is a local functional of $\uv$, i.e. it presents no coupling between displacement fields at different positions.  Taking the functional derivative with respect to $\uv(\rv)$ we arrive at the desired expression for the kinetic force density:  
\ber\label{Fk}
-\frac{\delta T_2[\uv]}{\delta u_\mu} 
&=&
\partial_\alpha[2T_{\nu\mu,0}u_{\nu\alpha}+T_{\nu\alpha,0}\partial_\mu u_\nu]-\frac{1}{4m}\partial_\nu\partial_\mu (n_0\partial_\nu \nabla\cdot\uv)\nn\\
&+&\frac{1}{4m}\partial_\nu \left\{ 2(\nabla^2 n_0) u_{\nu\mu}+(\partial_\nu n_0)\partial_\mu \nabla\cdot\uv+(\partial_\mu n_0)\partial_\nu \nabla\cdot\uv-2\partial_\mu\left[ (\partial_\alpha n_0)u_{\nu\alpha}\right]\right\}\,.
\eer

\subsection{Calculation of  $\delta E_2[\uv]/\delta\uv$ -- potential part}

To calculate the potential energy functional $W[\uv]$ we need the two-particle density matrix of the ground-state:
\be\label{defrho2}
\rho_2(\rv,\rv')\equiv\langle \psi_0|\hat \Psi^\dagger(\rv) \Psi^\dagger(\rv')\hat \Psi(\rv')\hat \Psi(\rv)|\psi_0\rangle\,.
\ee
For a system of electrons interacting via Coulomb interaction (charge $-e$) we have
\ber
W[\uv] =\frac{e^2}{2}\int d\rv\int d\rv' \frac{\rho_2(\rv,\rv')}{|\rv+\uv(\rv)-\rv'-\uv(\rv')|}\,.
\eer
Expanding to second order in $\uv$ we easily obtain
\be\label{W2-equation}
W_2[\uv] = -\frac{1}{2} \int d\rv \int d\rv' [u_\mu(\rv)-u_\mu(\rv')]K_{\mu\nu}(\rv,\rv')[u_\nu(\rv)-u_\nu (\rv')]
\ee
where
\be\label{CoulombKernel}
K_{\mu\nu}(\rv,\rv')=\rho_2(\rv,\rv')\partial_\mu\partial_\nu^\prime \frac{e^2}{|\rv-\rv'|}\,.
\ee
\end{widetext}
Finally, taking the functional derivative with respect to $\uv(\rv)$ we get
\be\label{deltaW2}
-\frac{\delta W_2[\uv]}{\delta u_\mu(\rv)} = \int d\rv' K_{\mu\nu}(\rv,\rv') [u_\nu(\rv)-u_\nu (\rv')]\,.
\ee
Thus, the inclusion of interactions transforms our equation of motion into an integro-differential equation.  Notice, however, that the interaction contribution vanishes if $\uv(\rv)$ is constant in space, as expected from the translational invariance of the interaction.

\subsection{Equation of motion for one-dimensional systems}

The formulas presented in the preceding subsections simplify dramatically in one-dimensional systems, where the displacement field has only one component, $u(x)$, the strain tensor reduces to the derivative of the displacement field $u'(x)$, and the equilibrium kinetic stress tensor reduces to a scalar
 \be\label{T01D}
 T_0(x)=\frac{1}{m}\left[\partial_x\partial_{x'} \rho(x,x')\big \vert_{x=x'}-\frac{n_0''}{4}\right]\,.
 \ee
Then the combination on the last line of Eq.~(\ref{Fk}) vanishes and we are left with the simpler expression
\be\label{Fk1D}
-\frac{\delta T_2[\uv]}{\delta u_\mu}  = (3T_0u')'-\frac{1}{4m}(n_0u'')''
\ee
where the primes denote derivatives with respect to $x$.  The complete equation of motion for one dimensional systems is thus
\ber\label{eom.1dimension}
mn_0\partial_t^2 u &=& -n_0u V_0^{\prime\prime}+(3T_0u')'-\frac{1}{4m}(n_0u'')''\nn\\&+&\int dx' K(x,x')[u(x)-u(x')]-n_0V_1^\prime\nn\,,\\
\eer
where $K(x,x')$ is given by the one-dimensional version of Eq.~(\ref{CoulombKernel}).
We will make use of this form of the equation of motion in the model applications presented below.

\section{Current-density functional approach}

Our discussion thus far has not relied on time-dependent current density functional theory, except on a very abstract level, i.e. as a basis for the statement that the stress tensor must be a functional of the current density.    The formulas presented in the last two subsections relied on the knowledge of the exact density matrices $\rho$ and $\rho_2$ of the many-body ground-state -- two  quantities that are amenable to treatment by powerful numerical techniques (e.g. the quantum Monte Carlo method) which have little in common with DFT.   Before proceeding, we wish to clarify how the time-dependent CDFT can help us in more concrete ways when  the exact $\rho$ and $\rho_2$ are {\it not} known, which is by far the most common case.  

One of the main ideas of  TDCDFT  is that the current and density evolutions of the interacting many-body system can be simulated in a non-interacting many-body system subject  to an effective time-dependent vector potential which includes the Hartree electrostatic potential and dynamical exchange-correlation (xc) effects.  This non-interacting system is known as the ``Kohn-Sham reference system" and its ground-state density coincides with the exact ground-state density of the interacting system, i.e. $n_0(\rv)$.  The potential that produces this exact ground-state density in the Kohn-Sham reference system is known as the static Kohn-Sham potential and is usually written as
\be
V_{s,0}(\rv)=V_0(\rv)+V_{H,0}(\rv)+V_{xc,0}(\rv)\,,
\ee  
where $V_{H,0}$ and $V_{xc,0}$ are, respectively, the Hartree potential and the xc potential of the ground-state.  These static potentials should not to be confused with the additional dynamical Hartree and xc potentials, which appear when the system is not in equilibrium. 
  
The idea is now to apply our continuum mechanics formulation directly to the Kohn-Sham reference system.  There is a small technical problem in doing this, namely the time-dependent xc vector potential $\Av_{xc}(\rv,t)$ that acts on the Kohn-Sham system has in general a transverse component, which cannot be represented as the gradient of a scalar potential.  Indeed, a complete representation of $\Av$  requires that we introduce both an electric field $\Ev_{xc}$  and a magnetic field $\Bv_{xc}$.  The inclusion of the xc magnetic field does not create any difficulties in principle (see Ref.~\onlinecite{tokatly2007}), and leads to the appearance of a Lorentz-force term in the equation of motion for the current.  But this Lorentz force term can be safely disregarded in the linear response approximation, because it has the form $\jv \times \Bv$ which is of second order in the deviation from equilibrium.  Thus we can take into account dynamical xc effects simply by adding the force $-e\Ev_{xc}$ to the driving force $-n_0\nablabold V_1$ on the right hand side of Eq.~(\ref{eom.simple}).  
All this considered, our equation of motion takes the form
\ber\label{eom.CDFT.final}
&&m\partial_t^2 \uv (\rv,t) + (\uv\cdot\nablabold) \nablabold V_{s,0}= -\frac{1}{n_0} \frac{\delta T_{s2}[\uv]}{\delta \uv(\rv)}\nn\\
&-&\nablabold [V_1(\rv,t)+V_{H,1}(\rv,t)]-e\Ev_{xc,1}(\rv,t)\,,
\eer     
where $V_{H,1}$ is the first-order term in the expansion of the time-dependent Hartree potential, and $\Ev_{xc,1}$ is the first order term in the expansion of $\Ev_{xc}$ in powers of $\uv(\rv)$.
The non-interacting kinetic force density $-\delta T_{s2}[\uv]/\delta \uv(\rv)$ is given by Eq.~(\ref{Fk}) in which, however, the equilibrium kinetic stress tensor $T_{\mu\nu,0}$ is replaced by the corresponding quantity for the Kohn-Sham reference system, i.e
\ber\label{Tsmunu0}
T^s_{\mu\nu,0}&=&\frac{1}{2m}\left(\partial_\mu\partial_\nu^\prime+\partial_\nu\partial_\mu^\prime\right) \sum_\ell \psi_{\ell}^*(\rv)\psi_{\ell}(\rv')\Big\vert_{\rv=\rv'} \nn\\&-& \frac{1}{4m}\nabla^2 n_0\delta_{\mu\nu}
\eer
where $\psi_{\ell}(\rv)$ are Kohn-Sham orbitals for the ground-state and the sum over $\ell$ runs over the occupied orbitals.

Assuming that the  Kohn-Sham ground-state has been obtained by one of the available approximations for $V_{s,0}$, the remaining problem is to find a suitable approximate expression for $\Ev_{xc,1}$.  The ``natural" approximation, in the present context, would be the high-frequency approximation, which expresses $\Ev_{xc,1}$ as the functional derivative of the exchange-correlation energy functional with respect to $\uv$.  In practice, since the latter is not known, one has to rely on more or less uncontrolled approximations, such as the high-frequency limit of the local density approximation proposed in Refs.~\cite{VK,vuc,tokatly2005b,tokatlyLNP2006} -- see Eq. (117) of Ref.~\onlinecite{tokatly2005b} (a general discussion of these approximation can be found in Ref.~\onlinecite{UllTok2006}).  This approximation is local both in space and time and is obtained by applying an instantaneous geometric deformation to the homogeneous electron gas. In the second respect it is perfectly consistent with our elastic approximation for the noninteracting kinetic force, but we must keep in mind that the latter is fully nonlocal.  

Unfortunately, the local deformation approximation for the xc potential suffers, like all electron-gas based
approximations, from a serious defect:  it fails to cancel the unphysical self-interaction that is contained in the Hartree term.  This makes it unsuitable for the treatment of strongly correlated system, where the spurious self-interaction energy can be very large.  More accurate approximations\cite{PSTS} do not suffer from this defect, but are more difficult to implement.  Furthermore, such approximations would do little (apart from fixing the self-interaction problem)  to capture the physics of strongly correlated electrons.  Alternatively, one could use functionals explicitly designed for electronic systems in the strong coupling limit, such as the ones developed by Seidl, Perdew and Kurth~\cite{Seidl00} and, more recently by Seidl, Gori-Giorgi, and Savin~\cite{Seidl07}.  

\section{The calculation of excitation energies}
\subsection{The eigenvalue problem}

An immediate application of our equation of motion is the calculation of excitation energies.\cite{Casida95,Gross96}  To this end we turn off the external potential $V_1$ and consider the homogeneous equation
\be
m n_0 (\rv)\partial_t^2 \uv(\rv,t)=-\frac{\delta E_2[\uv]}{\delta \uv(\rv,t)}\,.
\ee
Fourier-transforming with respect to time and carrying out the indicated expansion of the energy to second order in $\uv$ we get
\be\label{LinearResponse0}
m\omega^2 n_0(\rv)u_\mu(\rv,\omega)=\int d\rv' \left.\frac{\delta^2 E[\uv]}{\delta u_\mu(\rv)\delta u_\nu(\rv')}\right\vert_{\uv=0} u_\nu(\rv',\omega)\,.
\ee
Although this expression is not the most useful in practice, it does bring forth some important features of the problem. First, because the kernel of the integral equation is a symmetric second functional derivative, we are in the presence of an essentially hermitian eigenvalue problem.  More precisely, the problem is hermitian with respect to a scalar product with weight $n_0(\rv)$, i.e. defined as 
\be\label{defscalar}
(f,g)\equiv \int d\rv n_0(\rv) f(\rv) g(\rv)\,,
\ee
where $f$ and $g$ are arbitrary functions.  This can be seen by rewriting Eq.~(\ref{LinearResponse0})  as an equation for $\tilde \uv \equiv \sqrt{n_0(\rv)}\uv(\rv)$ and noting that this equation has the form of a standard eigenvalue problem with a symmetric kernel:
\ber\label{LinearResponse-second}
&&\int d\rv' \frac{1} {\sqrt{mn_0(\rv)}} \left.\frac{\delta^2 E[\uv]}{\delta u_\mu(\rv)\delta u_\nu(\rv')}\right\vert_{\uv=0} \frac{1} {\sqrt{mn_0(\rv')}} \tilde u_\nu(\rv',\omega)\nn\\&& =\omega^2 \tilde u_\mu(\rv,\omega)\,.
\eer

 This means that all the eigenvalues will be real, and eigenfunctions $\tilde \uv$ corresponding to different eigenvalues are orthogonal with respect to the ordinary scalar product. It follows that the original eigenfunctions $\uv$ are orthogonal with respect to the scalar product defined by Eq.~(\ref{defscalar}).  Second, the kernel of the integral equation is positive definite, because $E[\uv]$ has an absolute minimum at $\uv=0$, which corresponds to the ground-state energy.  For this reason, it is guaranteed that all the eigenvalues are positive.  The square roots of these eigenvalues are the approximate excitation energies of the system, starting from the ground-state. The eigenfunctions also have a simple interpretation as approximate matrix elements of the current density operator between the ground-state and the excited state under consideration, divided by the ground-state density.  We will show this more clearly in the next section.

\subsection{Derivation from linear response theory}

Additional insight into the significance of the eigenvalue problem for the excitation energies is obtained by deriving the equation of motion  directly from the linear response of the current density to an external vector potential in the high-frequency regime. To this end we write
\be\label{LinearResponse1}
j_\mu(\rv,\omega) = \int d\rv' \chi_{\mu\nu}(\rv,\rv',\omega)A_{\nu,1}(\rv',\omega)\,,
\ee
where $\chi_{\mu\nu}(\rv,\rv',\omega)$ is the current-current response function, and notice that, at high frequency, this function has the well-known expansion\cite{gvbook}
\be
\chi_{\mu\nu}(\rv,\rv',\omega)=\frac{n_0(\rv)}{m}\delta(\rv-\rv')\delta_{\mu\nu}+\frac{M_{\mu\nu}(\rv,\rv')}{m^2\omega^2}\,,
\ee
where the first term (diamagnetic term) is frequency-independent and
\be\label{FirstSpectralMoment}
M_{\mu\nu}(\rv,\rv') \equiv -m^2 \langle \psi_0 \vert [[\hat H_0,\hat j_\mu(\rv)],\hat j_\nu(\rv')]\vert\psi_0\rangle\,,
\ee
is the first spectral moment of the current-current  response function. ($[\hat A,\hat B]$ denotes the commutator of two operators $\hat A$ and $\hat B$, and $\psi_0$ is the undeformed ground-state wave function.)  Now, replacing 
$\jv(\rv,\omega)=-i\omega n_0(\rv)\uv(\rv,\omega)$ and $\Av_1(\rv,\omega) = \frac{\nablabold V_1(\rv,\omega)}{i\omega}$, and solving for $\nablabold V_1$ to leading order in $1/\omega^2$ we obtain
\be\label{eom.last}
\partial_\mu V_1 = m \omega^2 u_\mu -\frac {1}{n_0}\int d\rv' M_{\mu\nu}(\rv,\rv') u_\nu(\rv')\,.
\ee
This is equivalent to our equation of motion~(\ref{eom.exact2}) if and only if 
\be\label{Moment-Energy}
 M_{\mu\nu}(\rv,\rv')= \left.\frac{\delta^2 E[\uv]}{\delta u_\mu(\rv)\delta u_\nu(\rv')}\right\vert_{\uv=0}\,.
 \ee
To show that this is indeed the case we observe that the deformed ground-state wave function can be expanded as 
\be
\psi_0[\uv]=\psi_0+\psi_0^{(1)}+\psi_0^{(2)}+...
\ee
where $\psi_0$ is the undeformed ground-state wave function and $\psi_0^{(1)}$, $\psi_0^{(2)}$ are corrections of first and second order in $\uv$ respectively.  The various corrections are not mutually independent.  If $\psi_0[\uv]$ is normalized to a constant independent of $\uv$, then we must have
\be
\langle\psi_0|\psi_0^{(2)}\rangle+\langle\psi_0^{(2)}|\psi_0\rangle=-\langle\psi_0^{(1)}|\psi_0^{(1)}\rangle\,.
\ee
Taking this into account it is easy to verify that the second order correction to the energy is
\be\label{e2psi1}
E_2[\uv] = \langle\psi_0^{(1)}|\hat H_0-E_0|\psi_0^{(1)}\rangle\,,
\ee
where $E_0$ is the ground-state energy.  Finally, we observe that the first-order correction to the ground-state wave function is given by
\be\label{psi1}
|\psi_0^{(1)}\rangle = -im\int d\rv \hat j(\rv)\cdot \uv(\rv)|\psi_0\rangle
\ee
where we have used the fact that the momentum density operator $m\hat \jv(\rv)$ is the generator of a local translation of all the particles in an infinitesimal volume located at $\rv$.  Thus, the operator on the right hand side of Eq.~(\ref{psi1})  performs different translations by vectors $\uv(\rv)$ at different points in space, i.e. precisely deforms the ground-state according to the displacement field $\uv(\rv)$.  Substituting the above expression for $|\psi_0^{(1)}\rangle$ into Eq.~(\ref{e2psi1})  for $E_2[\uv]$  one can easily verify that
\be
E_2[\uv]=\frac{1}{2}\int  d\rv \int d\rv' u_\mu(\rv) M_{\mu\nu}(\rv,\rv') u_\nu(\rv')\,,
\ee
with $M_{\mu\nu}(\rv,\rv')$  given by Eq.~(\ref{FirstSpectralMoment}).   This establishes the validity of Eq.~(\ref{Moment-Energy}).

\subsection{Eigenfunctions and sum rule}

The spectral representation of the kernel of our equation of motion gives additional insight into the nature of our approximation and shows clearly where things can go wrong.  Namely we can write
\ber
M_{\mu\nu}(\rv,\rv')&=&m^2\sum_n \omega_{n0}\left\{[ j_{\mu}]_{0n}(\rv)[j_{\nu}]_{n0}(\rv')\right.\nn\\&+&\left. [j_{\nu}]_{0n}(\rv')[j_{\mu}]_{n0}(\rv)\right\}\,,
\eer
where $\omega_{n0}$ are the exact excitation energies of the system from the ground-state (0) to the n-th excited state, and $[j_{\mu}]_{n0}(\rv)\equiv\langle n|\hat j_\mu(\rv)|0\rangle$ are the matrix elements of the current density operator between the corresponding states.  An exact linear equation for the excitations would have to give $\pm\omega_{n0}$ as excitation energies and $n_0^{-1}(\rv)[j_{\mu}]_{0n}(\rv)$ ($=n_0^{-1}(\rv)[j_{\mu}]_{n0}^*(\rv)$)  as the corresponding eigenfunctions.  

In general this will not be the case.  However, for the special case of a one-particle system, in the absence of a magnetic field,  we can show rather easily that  $[j_{\mu}]_{0n}(\rv)= - [j_{\mu}]_{n0}(\rv)$ and furthermore that $n_0^{-1}(\rv)[j_{\mu}]_{0n}(\rv)$  is indeed an eigenfunction of the operator $n_0^{-1}(\rv) M_{\mu\nu}(\rv,\rv')$ with eigenvalue $m\omega_{n0}^2$.  This follows from the orthonormality relation
\be\label{1P-orthonormality}
\frac{2m}{\sqrt{\omega_{n0}\omega_{k0}}}\int d\rv \frac{[j_{\nu}]_{0n}(\rv)[j_{\nu}]_{k0}(\rv)}{n_0(\rv)} = \delta_{nk}\,,
\ee
which is valid for one-particle systems (in the absence of a magnetic field) and is proved in Appendix D.  Then the eigenvalues of the equation of motion~(\ref{eom.last}) with $V_1=0$ are $\omega=\pm\omega_{n0}$ as they should be.   This result is perfectly consistent with our previous observation that the ``elastic approximation" is not an approximation at all when it comes to one-electron system, due to the lack of retardation effects in such systems.  

In  general, in a many-particle system the matrix elements of the current density operator between the ground state and different excited states are not necessarily orthogonal.  Indeed, we will see that two completely different excited states can produce the same eigenfunction for the displacement field, up to a proportionality constant.    The reason why this can happen is that the {\it exact} equation of motion for the displacement field of a many-body system is not an eigenvalue problem (even though it is linear) due to the frequency dependence of the kernel.  As a result, the normalization of the solutions becomes relevant:  two ``eigenfunctions" that differ by a mere proportionality constant can result in different excitation energies when the kernel of the linear equation is itself  energy-dependent.  In such cases, the elastic approximation will fail to resolve the different excitation energies, replacing them by a single excitation energy at an ``average" value.

In spite of this shortcoming, an exact sum rule can be established, which relates the exact eigenvalues $\omega_{\lambda}$ of the elastic eigenvalue problem  to the exact excitation energies $\omega_{n0}$.  The sum rule reads:
\be\label{SumRule}
\omega_\lambda^2 = \sum_n f^\lambda_n \omega_{\lambda0}^2
\ee
where the ``oscillator strengths"
\be\label{OscillatorStrength}
f^\lambda_n= \frac {2 m \left\vert \int d\rv  \jv_{0n}(\rv) \cdot \uv_\lambda (\rv)\right\vert^2}{\omega_{n0}}\,,
\ee 
and where $\uv_\lambda(\rv)$ is the solution of the elastic eigenvalue problem normalized with respect to the scalar product~(\ref{defscalar}), $\omega_\lambda$ is its eigenvalue, and the sum runs over all the exact eigenfunctions.  Further, the oscillator strengths satisfy the sum rule
\be\label{OscillatorStrengthSumRule}
\sum_n f^\lambda_n=1
\ee
for all $\lambda$.  The proof of these results is presented in Appendix E. 

From this vantage point we see that the elastic approximation is the extension of the well-known collective approximation\cite{Collective} of the homogeneous electron gas to inhomogeneous systems.  Each eigenvalue $\omega_\lambda$ of the elastic equation of motion is a weighted average of exact excitation energies, with a weight controlled by the overlap of the exact current matrix element with the eigenmode $\uv_\lambda(\rv)$.  Since  $\uv_\lambda(\rv)$ form an orthogonal basis in the space of displacements, we can say that $\omega_\lambda$ represents the average energy of exact excitations in the ``direction" $\lambda$. Thus,  the full excitation spectrum is replaced by a set of spectral lines, one for each orthogonal direction in displacement space, and each one carrying the entire and exact spectral weight for that particular direction.

\section{Model applications}
For orientation we now examine the application of our theory to a few simple models.

\subsection{Homogeneous electron gas}

In a homogeneous electron gas the ground-state density $n_0=n$ is independent of position.   The equilibrium kinetic stress tensor has a constant value
\be
 T_{\mu\nu,0} = \frac{2}{3} nt(n)\delta_{\mu\nu}\,,
 \ee
 where $t(n)$ is the kinetic energy per particle.    The  two particle density matrix $\rho_2(\rv,\rv')$ is a function of $|\rv-\rv'|$.  In such a homogeneous system the displacement eigenfunctions are simply plane waves  
 \be
 \uv(\rv,\omega) = \tilde \uv(\qv,\omega) e^{i(\qv\cdot\rv - \omega t)}
 \ee
 characterized by a wave vector $\qv$.  Of these there are two kinds:  longitudinal, in which $\tilde \uv$ is parallel to $\qv$, and transverse, in which $\tilde \uv$ is perpendicular to $\qv$.   The expression~(\ref{Fk})  for the kinetic force density reduces to 
\ber
-\frac{\delta T_2[\uv]}{\delta\uv} &=& - \frac{2}{3}nt(n)[2\qv (\qv\cdot \tilde \uv) +q^2 \tilde\uv]+\frac{nq^2}{4m}\qv (\qv\cdot\tilde\uv)\,.\nn\\
\eer
The force density from potential energy, Eq.~(\ref{deltaW2}), is given by
\ber
-\frac{\delta W_2[\uv]}{\delta\uv_\mu} &=&  [K_{\mu\nu}({\bf 0})-K_{\mu\nu}(\qv)]\tilde u_\nu\,,
\eer
where
\be
K_{\mu\nu}(\qv)=-\int \frac{d\qv'}{(2\pi)^3}\rho_2(\qv-\qv') q'_\mu q'_\nu v(q')\,,
\ee
where $v(q)=4\pi e^2/q^2$ is the Fourier transform of the Coulomb potential in three dimensions, and $\rho_2(\qv)$ -- the Fourier transform of $\rho_2(\rv-\rv')$  -- is related to the static structure factor $S(q)$ in the following manner:\cite{gvbook} 
\be
\rho_2(\qv)=n[S(\qv)-1]\,.
\ee
Thus, we get
\ber
-\frac{\delta W_2[\uv]}{\delta\uv} = -n\int d\qv' [S(\qv-\qv')-S(\qv')]v(q')\qv'[\qv'\cdot \tilde\uv (\qv)]\,.\nn\\
\eer
Finally, in order to take into account the neutralizing background of positive charge (required for the stability of the electron gas), we add the external potential
\be
V_0(\rv)=\frac{m}{2}\omega_p^2 (\rv \cdot \hat \qv)^2\,,
\ee
where $\omega_p^2=4\pi n e^2/m$ is the square of the plasmon frequency and $\hat \qv$ is the unit vector in the direction of $\qv$.  Notice that this potential is assumed to vary only in the direction of $\hat \qv$, because it is only in this direction that the displacement generates boundary charges: the system remains perfectly homogeneous in the direction perpendicular to $\qv$.\footnote{Alternatively, we could set $V_0=0$ and include the $\qv=0$ singularity in the structure factor: $\rho_2(\qv)=n^2\delta(\qv)+ n[S(\qv)-1]$.}   The corresponding force  in the equation of motion~(\ref{eom.simple}) is
\be
\uv \cdot \nablabold\partial_\mu V_0 = m\omega_p^2 (\hat \qv \cdot \uv)  \hat q_\mu\,.
\ee  

\begin{widetext}
Putting everything together, the equation of motion takes the form
\ber
m\omega^2 \tilde \uv = \frac{2}{3}t(n) \left[2\qv(\qv \cdot \tilde \uv)+q^2\tilde \uv\right]+ \frac{q^2}{4m}\qv (\qv\cdot\tilde\uv) + m\omega_p^2 (\hat \qv \cdot \tilde \uv)  \hat \qv 
+\frac{m\omega_p^2}{n}\int \frac{d\qv'}{(2\pi)^3} [S(\qv-\qv')-S(\qv')]\hat \qv'[\hat \qv'\cdot \tilde \uv (\qv)]\,,\nn\\
\eer
This can be further decoupled into longitudinal and transverse components denoted by $\tilde u_L$ and $\tilde u_T$ respectively.  The corresponding eigenvalues are
\be\label{omegaL}
\omega^2_L(q) =   \omega_p^2+2t(n)\frac{q^2}{m}+\frac{q^4}{4m^2}+\frac{\omega_p^2}{n}\int \frac{d\qv'}{(2\pi)^3}\left(\hat \qv \cdot \hat \qv'\right)^2 [S(\qv-\qv')-S(\qv')]\,,
\ee
and
\be\label{omegaT}
\omega^2_T(q) =   \frac{2}{3}t(n)\frac{q^2}{m}+\frac{\omega_p^2}{2n}\int \frac{d\qv'}{(2\pi)^3} \left(\hat \qv \times \hat \qv'\right)^2 [S(\qv-\qv')-S(\qv')]\,.
\ee
\end{widetext}
Of course, this is exactly what one would have obtained by assuming that the spectrum of the current-current response function consist of a single $\delta$-function peak at $\omega_L$ or $\omega_T$ and requiring satisfaction of the first moment sum rule (see Ref. \onlinecite{gvbook}, Eq. 3.191 and Ref. \onlinecite{Pathak73}).

\begin{figure}
\begin{center}
\includegraphics[width=0.9\linewidth]{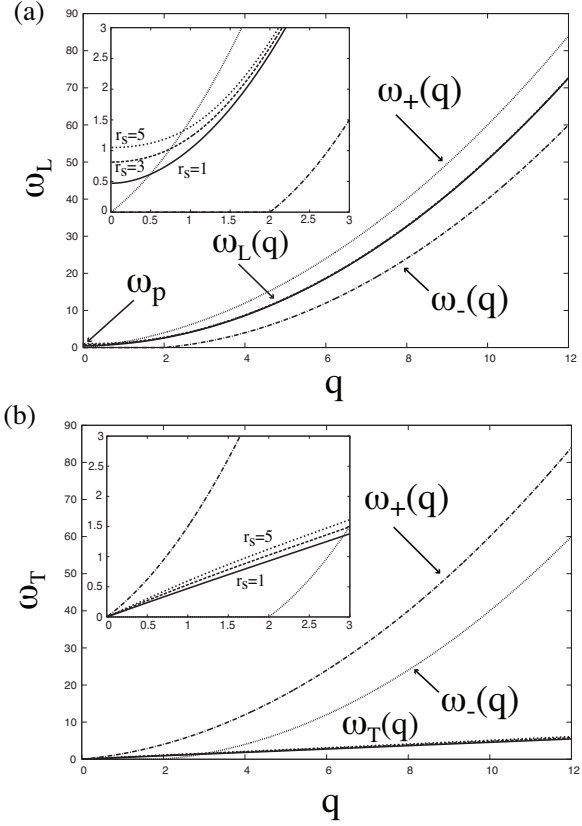}
\caption{(Color online)  Longitudinal (a) and transverse (b) modes for a homogeneous electron gas at $r_s=1,3,5$. Wave vector $q$ is in units of $k_F$ and frequency is in units of $2E_F$. The curves labelled $\omega_+(q)$ and $\omega_-(q)$ are the boundaries of the electron-hole continuum.\cite{gvbook}. Single particle excitations exist for $\omega_-(q)<\omega<\omega_+(q)$, whereas multiparticle excitations are distributed all over the plane.  The elastic approximation replaces the exact spectrum by the two branches $\omega_L(q)$ and $\omega_T(q)$, which carry the entire spectral weight.  Notice that the $r_s$-dependence is barely discernible on a large $q$ scale, but becomes clearly visible at smaller $q$ as shown in the insets.} 
\label{OLTK}
\end{center}
\end{figure}

In Fig.~(\ref{OLTK}) we plot the excitation spectrum of the homogeneous electron gas calculated from Eqs.~(\ref{omegaL}) and (\ref{omegaT}).  We have used the static structure factor calculated in Ref.~\onlinecite{PGG00} by the quantum Monte Carlo method, and the kinetic energy  has been computed 
from the parametrized correlation energy of Appendix B of Ref. \onlinecite{PGG00}, using the virial theorem.
In the longitudinal channel, the exact spectrum is dominated at small $q$ by the plasmon and at large $q$ by free particle excitations (energy $q^2/2m$).  There are also also electron-hole pair excitations at lower $q$ and $\omega$, as well as multiple electron-hole pair excitations all over the plane.  The elastic approximation replaces this complex spectrum by a single branch of longitudinal excitations which has the correct spectral moment. In particular, we get the correct dispersion of the plasmon at small $q$ and the correct free particle behavior at large $q$.  In the transverse channel the plasmon and the high-$q$ free particle excitations are absent.  The exact transverse spectrum consists primarily of low-energy electron-hole pair excitations. The current vector, $\kv+\qv/2$, where $\kv$ and $\kv+\qv$ are the momenta of the hole and the electron, respectively, is essentially perpendicular to $\qv$ when $\kv$ and $\kv+\qv$ both lie near the Fermi surface.   On the other hand, high-energy excitations, with energy ~$q^2/2m$ are essentially longitudinal because the current vector  is essentially parallel to $\qv$ when $\qv$ is much larger than the Fermi momentum.  This is consistent with the fact that the frequency of our  transverse collective mode  $\omega_T(q)$ grows linearly with $q$ at large $q$.

\subsection{Linear harmonic oscillator and hydrogen atom}

In order to demonstrate the exactness of the formulation for one-electron systems let us now consider the canonical examples  of the one-dimensional harmonic oscillator and the hydrogen atom.   

For a harmonic oscillator of natural frequency $\omega_0$, external potential $V_0(x)=m\omega_0^2x^2/2$, and equilibrium density $n_0(x) = \frac{e^{-x^2/\ell^2}}{\sqrt{\pi} \ell}$, where $\ell \equiv (m\omega_0)^{-1/2}$,  the equation of motion~(\ref{eom.1dimension})  reduces to
\be\label{HOequation}
\frac{1}{4}u'''' -xu''' + (x^2 - 2)u''+ 3x u' - \frac{\omega^2-\omega_0^2}{\omega_0^2} u = 0\,.
\ee
Solving the eigenvalue problem with the boundary condition of $n_0^{1/2}(x) u(x) \rightarrow 0$ as $|x| \rightarrow \infty$, we obtain the exact excitation spectra $\omega_n = \pm n\Omega$,
where $n = 1, 2, ..$. The corresponding eigenfunctions are 
\be
u_n(x) \propto H_{n-1}(x)
\ee
which are mutually orthogonal with respect to the scalar product~(\ref{defscalar}).  These are indeed proportional to the matrix elements of the current density operator between the ground-state and the n-th excited state.

A similar calculation can be done for hydrogen-like atoms of atomic number $Z$. 
Focusing for simplicity on excitations of spherical symmetry, we introduce a radial displacement field $u_r(r)$ which depends only on the radial coordinate $r$.   Then $u_r$ satisfies the equation
\begin{eqnarray}
\frac{1}{4}u_r'''' - \left(1 - \frac{1}{r}\right)u_r''' +
\left(1 - \frac{2}{r} - \frac{1}{r^2}\right)u_r'' \nn \\
+ \frac{3}{r^2} u_r'
- \left(\frac{2}{r^3} + \frac{\omega^2}{Z^4}\right)u_r = 0\,,
\end{eqnarray}
where the primes now denote derivatives with respect to $r$.
Solving the eigenvalue problem with boundary condition $n_0^{1/2}(r)u_r(r) \to 0$ for $r\to\infty$ yields the correct excitation energies
$\omega_n =(Z^2/2)(1 - 1/n^2)$ ($n = 1, 2, ..$).   The corresponding eigenfunctions are given by Laguerre polynomials:
\be
u_n(r)\propto L_{n-2}^2\left(\frac{2r}{n}\right)\,.
\ee

\subsection{Two-electron systems}

As a final example let us consider the case of two electrons
repelling each other with the ``soft" Coulomb potential
$\frac{e^2}{\sqrt{(x_1-x_2)^2+a^2}}$ in a one-dimensional parabolic trap of natural frequency $\omega_0$
(the
cutoff $a>0$ serves to eliminate the pathological behavior of the interaction at
$x_1=x_2$).   This is a model that can be solved numerically
thanks to the separation of center of mass and relative variable,
and analytically in the limit of strong correlation.  
The hamiltonian is
\be
\label{H2e}
\hat H_0 = \frac{P^2}{4m}+m\omega_0^2 X^2+\frac{p^2}{m}+\frac{m}{4}\omega_0^2 x^2+\frac{e^2}{\sqrt{x^2+a^2}}\,,
\ee
where $X=\frac{x_1+x_2}{2}$ and $P=p_1+p_2$ are, respectively, the coordinate and the momentum of the center of mass and $x=x_1-x_2$, $p=\frac{p_1-p_2}{2}$ are the coordinate and the momentum in the relative channel.  Notice that for a fixed strength $e^2$ of the interaction we can go from the weakly correlated to  the strongly correlated regime by varying the value of $\omega_0$:  $\omega_0 \to \infty$ corresponds to the non-interacting limit, and $\omega_0 \to 0$ to the strongly correlated limit.

The ground-state wave function is
\be
\Psi_0(x_1,x_2)=\psi_0(X)\phi_0(x)\,,
\ee
where $\psi_0(X)$ and $\phi_0(x)$ are, respectively, the ground-state wave functions of the center of mass and of the relative hamiltonian.
The general excited state is
\be
\Psi_{nm}(x_1,x_2)=\psi_n(X)\phi_m(x)
\ee
where $\psi_n(X)$ is the wave function  of the n-th excited state of the center of mass hamiltonian and $\phi_m(x)$ is the wave function of the $m$-th excited state of the relative hamiltonian.

The ground-state of the system is a spin singlet  ($S=0$) and for this reason in the following we consider only singlet states, which are connected to the ground-state by the current density operator.  The relative wave function for such states is symmetric: $\phi_m(-x)=\phi_m(x)$.  This wave function has $2m$ nodes, of which $m$ with $x>0$ and $m$ (symmetrically placed) with $x<0$.  The center of mass wave function, 
\be
\psi_n(X)= H_{n}\left(\frac{X}{\ell_{cm}}\right) e^{-X^2/2\ell_{cm}^2}\,,
\ee
 can be either symmetric or antisymmetric, depending on the parity of $n$, and has $n$ nodes.   Here $\ell_{cm}=\sqrt{\hbar/2m\omega_0}$.  The ground-state has $n=0,m=0$ and all the other states are characterized by positive values of the integers $n$ and $m$.

In the non interacting limit ($\omega_0 \to \infty$, or $e^2 \to 0$) the relative wave function is
\be
\phi_m(x)= H_{2m}\left(\frac{x}{\ell_{0}}\right) e^{-x^2/2\ell_{r}^2}\,,
\ee
where  $\ell_{0}=\sqrt{2\hbar/m\omega_0}$. 
The ground-state density is a gaussian centered at the origin.   
The excitation energies, expressed in units of $\omega_0$, are sums of the excitation energies of two identical harmonic oscillators:
\be\label{EigenvaluesWeak}
\lim_{\omega_0 \to \infty} \frac{E_{nm}}{\omega_0}= n+2m
\ee
with $n$, $m$ non-negative integers.    The degeneracy of the excited states is the number of integers less or equal $n+2m$ with the same parity as $n+2m$, i.e. 
\be\label{DegeneracyWeak}
D_{nm}= 1+\left[\frac{n+2m}{2}\right]\,,
\ee   
where $[y]$ denotes the integer part of $y$.  The displacement field associated with the $(n,m)$ excitation is
\be\label{DisplacementFieldsWeak}
u_{nm}(x)\propto H_{n+2m-1}(x/\lambda_0)\,,
\ee
where $\lambda_0=\sqrt{\ell_{cm}^2+(\ell_{0}/2)^2}=\sqrt{\hbar/m\omega_0}$.   The parity is $(-1)^{n+2m-1}$ and the number of nodes in $n+2m-1$.    

\begin{figure}
\begin{center}
\includegraphics[width=0.9\linewidth]{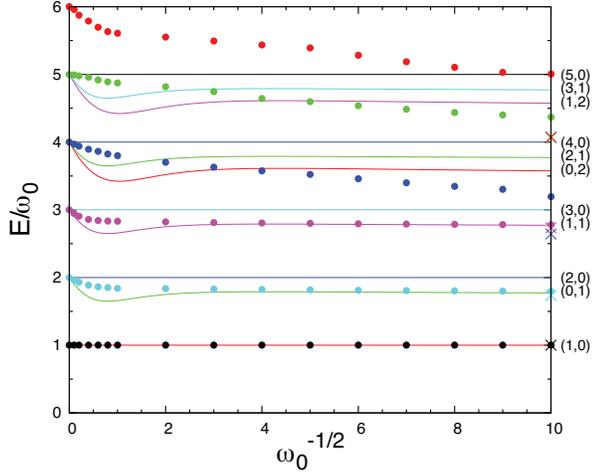}
\caption{(Color online)  Evolution of the excitation energies for two electrons in a one-dimensional harmonic trap. Solid lines denote exact excitation energies, labelled by $(n,m)$ as explained in the text.  Solid dots indicate the calculated eigenvalues of the QCM equation of motion.  Crosses on the right denote the strong-correlation limit of the eigenvalues, given by Eq.~(\ref{ExactEigenvalues}).}   
\label{EvolutionOfLevels}
\end{center}
\end{figure}

The situation is quite different in the strongly correlated limit  ($\omega_0 \to 0$, or $e^2 \to \infty$). The relative Hamiltonian reduces to a harmonic oscillator of frequency $\omega_0 \sqrt{3}$ with equilibrium distance $x_0 =(2e^2/m\omega_0^2)^{1/3}$ (we assume $a \ll x_0$). The ground-state wave function for the relative motion is a symmetric linear combination of two Gaussians of width $\ell_{\infty}=(2\hbar/\sqrt{3}m\omega_0)^{1/2}$ centered at $x_1-x_2 = \pm x_0$. The corresponding ground state density $n_0(x)$ consists of two Gaussian peaks of the width $\lambda_{\infty}=\sqrt{\ell_{cm}^2+(\ell_{\infty}/2)^2}=[\hbar (1+\sqrt{3})/2\sqrt{3}m\omega_0]^{1/2}$ centered at $x = \pm x_0/2$.  The excitation spectrum has the form
\be\label{EigenvaluesStrong}
\lim_{\omega_0 \to 0} \frac{E_{nm}}{\omega_0}=n+m\sqrt{3}\,,
\ee
and the degeneracy is completely removed.  The displacement field is analytically found to be
\ber\label{DisplacementFieldsStrong}
u_{nm}(x) &\propto&  H_{n+m-1}\left(\frac{x-x_0/2}{\lambda_{\infty}}\right)\theta(x)\nn\\
&+&(-1)^m H_{n+m-1}\left(\frac{x+x_0/2}{\lambda_{\infty}}\right)\theta(-x)\,.
\eer
The parity is $(-1)^{n-1}$ and the number of nodes is $2(n+m-1)+{\rm mod}(n-1,2)$, where mod$(n-1,2)\equiv n-1$ (mod 2).

\begin{figure}
\begin{center}
\includegraphics[width=0.8\linewidth]{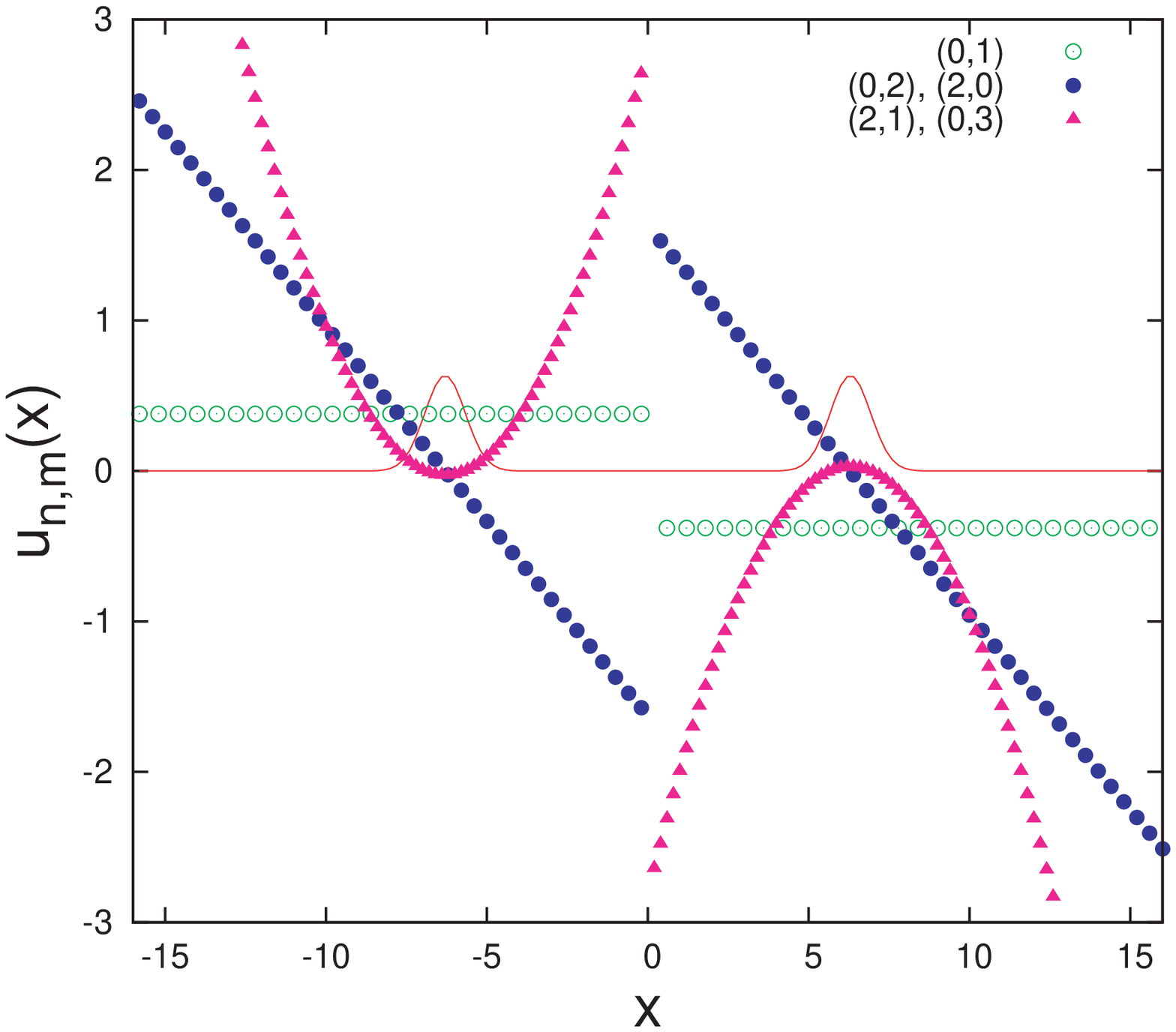}
\includegraphics[width=0.8\linewidth]{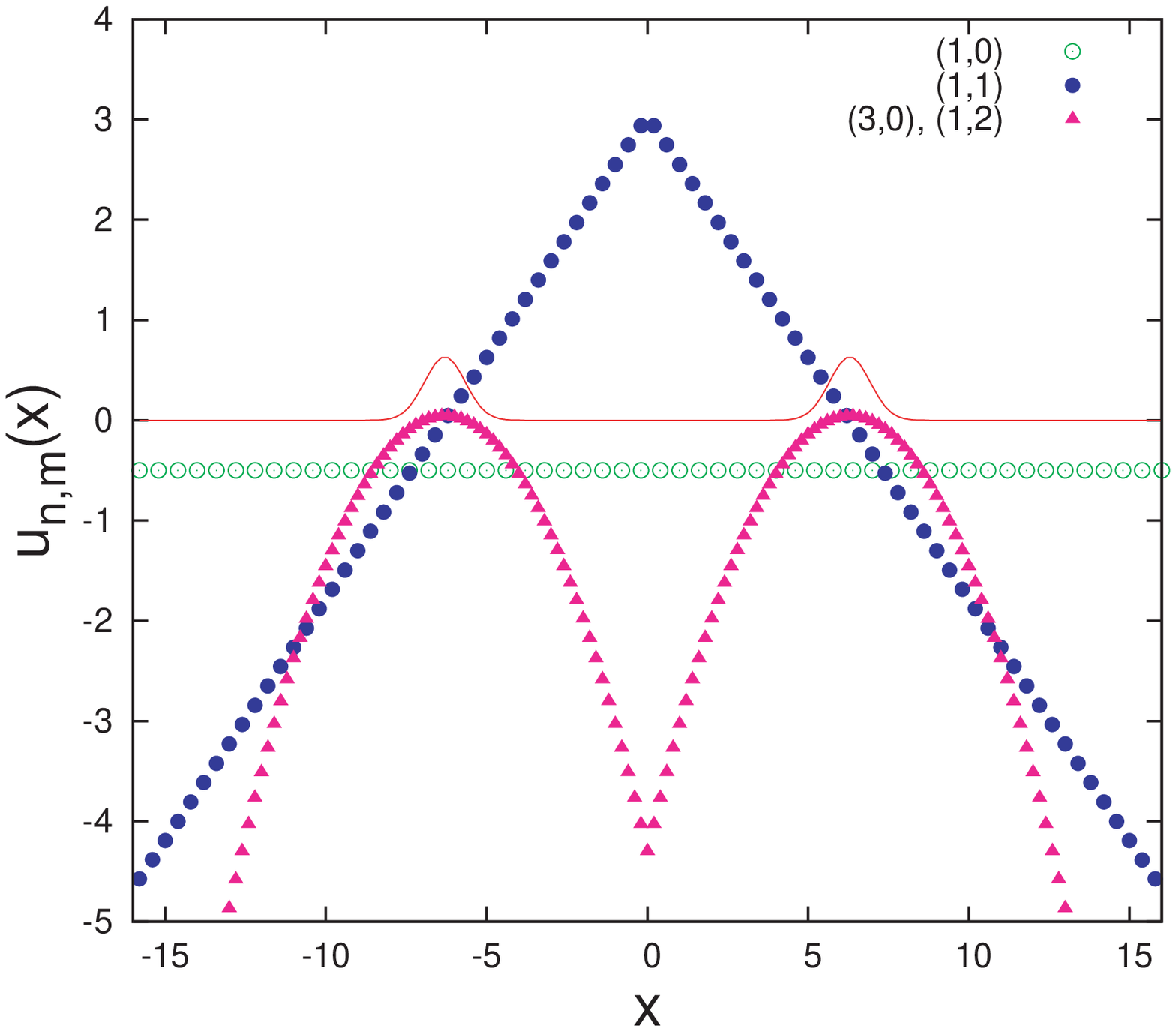}
\caption{(Color online)  Top panel: the displacement field $u_{nm}(x)$
for $(n,m)=(0,1), (0,2)$ and $(0,3)$ in the strong correlation limit.  Bottom panel: the same for $(n,m)=(1,1), (1,2)$ and $(2,1)$.  The thin solid lines represent the density profile.  The large value of the displacement field for $x\sim 0$  does not have a  physical significance since the density is exponentially small in that region. 
\label{fig:one}}
\label{DisplacementFields}
\end{center}
\end{figure}

The evolution of the lowest-lying energy levels with given value of the pair $n,m$ as a function of $\omega_0$ is shown by the solid lines in Fig.~(\ref{EvolutionOfLevels}).  Some of the displacement fields of the low-lying excitations in the strongly correlated regime (Eq.~(\ref{DisplacementFieldsStrong})) are shown in Fig.~(\ref{DisplacementFields}).

From these figures we see that the displacement field of the $(1,0)$ excitation, which corresponds to a rigid translation of the center of mass, is uniform in space, while the displacement field of the $(0,1)$ excitation, which corresponds to the classical breathing mode, changes sign around the origin.  The $(1,0)$ and $(0,1)$ excitations correspond to the classical phonon modes of a system of two localized particles.  The remaining excitations are quantum mechanical in character, as can be surmised from the fact that their displacement fields (in the strongly correlated regime) have significant variation over the regions where the density has peaks, i.e. the places where the particles would be classically localized.  These modes describe the dynamics of the wave function of the localized electrons.

Looking at the figures we observe that, {\it in the strongly correlated regime}, there are groups of excited states (e.g. $\{(0,2),(2,0)\}$;  $\{(3,0),(1,2)\}$;  $\{(2,1),(0,3)\}$; $\{(4,0),(0,2),(0,4)\}$,  such that all the excited states within one group produce the same displacement field, up to a normalization constant.  In general, states with a given value of $n+m$ and the same parity of $m$ have the same displacement field, but different energies.  Clearly, this is a feature of the exact solution that cannot be reproduced by any linear eigenvalue problem with a frequency-independent kernel.

The phenomenon of different excited states producing the same displacement field occurs also in the non-interacting limit: all the states with the same value of $n+2m$ (e.g. $(0,1)$ and $(2,0)$)  have the same displacement field.  But, in this case, the states with the same displacement field also have the same energy: therefore the noninteracting excitation energies can be accurately reproduced by a linear eigenvalue problem. 

Let us now see what our elastic equation of motion~(\ref{eom.1dimension}) predicts for this system.  
The kinetic part of the equilibrium stress tensor $T_0(x)$ works out to be
\begin{widetext}
\ber\label{eq:110}
T_0(x)=\frac{2}{m}\int dy\left\{ \left[\phi_0'(x-y)\psi_0\left(\frac{x+y}{2}\right)+ \frac{1}{2}\phi_0(x-y)\psi_0'\left(\frac{x+y}{2}\right)\right]^2
-\frac{1}{4}\partial^2_x\left[\phi_0^2(x-y)\psi_0^2\left(\frac{x+y}{2}\right) \right]\right\}
\eer
where  $\psi_0$ and $\phi_0$ are the ground-state wave function in the center of mass and relative channel respectively.
The interaction kernel $K(x,x')$ is given by
\be\label{eq:111}
K(x,x')=\frac{-2(x-x')^2+a^2}{[(x-x')^2+a^2]^{5/2}}\phi^2_0(x-x')\psi^2_0\left(\frac{x+x'}{2}\right)\,.
\ee
\end{widetext}
We now have all the input that is necessary to set up and solve the fourth-order integro-differential equation~(\ref{eom.1dimension}). 

In the limit of weak correlation ($\omega_0 \to \infty$) the eigenvalues of the integro-differential equation coincide with the exact (degenerate) excitation energies.   This is understandable, since in this limit the two electrons are decoupled and the excitation spectrum of the two-electron system coincides with that of a single electron starting from its own ground-state.  This spectrum, as we have seen,  is exactly reproduced by our equation of motion.  Unfortunately, this nice feature of the present model cannot be extrapolated to general systems.  If, for example, the system contains more than two electrons, then even in the non-interacting limit a generic excitation will entail the transition of a single electron from an occupied orbital that is {\it not} the ground-state orbital to an unoccupied one.  Such an excitation will {\it not} be described exactly by our method.

In the limit of strong correlation our integro-differential equation can be solved analytically, as shown in Appendix F.  The eigenfunctions can be classified as even  (+) or  odd (-)  and are given by
\ber \nonumber
u_{k,\pm}(x) &\propto&  H_{k}\left(\frac{x-x_0/2}{\lambda_{\infty}}\right)\theta(x)\\
&\pm& (-1)^k H_{k}\left(\frac{x+x_0/2}{\lambda_{\infty}}\right)\theta(-x)\,,
\label{ElasticDisplacementFieldsStrong}
\eer
where $k$ is a non-negative integer.  The number of nodes is $2k$ for even eigenfunctions, $2k+1$ for odd eigenfunctions.  The corresponding eigenvalues are given by
\ber\label{ExactEigenvalues}
\lim_{\omega_0\to 0} \frac{E_{k,\pm}}{\omega_0} &=& \left[2+3\sqrt{3}k +6k(k-1)(2-\sqrt{3})\right.\nn\\
&&\left.\mp (-1)^k  (2-\sqrt{3})^k\right]^{1/2}\,.
\eer
Notice that, within each symmetry sector (even or odd), the eigenvalues increase monotonically with increasing $k$.

In Table I, fourth and fifth column, we present a detailed comparison between the exact excitation energies and the eigenvalues of our equation of motion in the strong correlation limit.  For the sake of clarity, we list the excitations that produce even displacement fields and those that produce odd displacement fields separately.
\begin{table}
Even modes\\
~~\\
\begin{tabular}{|c|c|c|c|c|c|}
\hline 
$(n,m)$ &$E^0_{nm}$&${N^0}$& $E^\infty_{nm}$  &$E^\infty_{n+m-1,+}$ & ${N^\infty}$\\
\hline 
(1,0)&1.0&0& 1.0 & 1.0 &0\\ 
\hline
(1,1) &3.0&2&  2.732 & 2.732&2\\ 
\hline
(3,0) &3.0&2&  3.0 &3.942 & 4 \\ 
(1,2) &5.0&4&  4.464 &~ &4\\
\hline
(3,1) &5.0&4& 4.732&5.220 & 6 \\ 
(1,3)&7.0&6&  6.196 &~& 6\\
\hline
(5,0) &5.0&4& 5.0&6.486&8 \\
(3,2) &7.0&6& 6.464&~&8\\ 
(1,4)&9.0&8&  7.928 &~& 8\\
\hline
(5,1) &7.0&6&  6.732 & 7.755 &10\\ 
(3,3)&9.0&8&  8.196 & ~&10\\
(1,5)&7.0&6&  9.660 & ~&10\\
\hline 
\end{tabular}\\
~~\\
~~\\
Odd modes\\
~~\\
\begin{tabular}{|c|c|c|c|c|c|}
\hline 
$(n,m)$ &$E^0_{nm}$&${N^0}$& $E^\infty_{nm}$  &$E^\infty_{n+m-1,-}$ & ${N^\infty}$\\
\hline 
(0,1) &2.0&1&  1.732 & 1.732 &1\\ 
\hline
(2,0) &2.0&1&  2.0 &2.632& 3 \\ 
(0,2) &4.0&3&  3.464&~ &3\\
\hline
(2,1) &4.0&3& 3.732&3.960 & 5 \\ 
(0,3)&6.0&5& 5.196 &~& 5\\
\hline
(4,0) &4.0&3& 4.0&5.217& 7 \\ 
(2,2) &6.0&5& 5.464&~& 7 \\ 
(0,4)&8.0&7&  6.928 &~& 7\\
\hline
(4,1) &6.0&5&  5.732 & 6.487& 9\\ 
(2,3)&8.0&7&  7.196 & ~&9\\
(0,5)&10.0&9&  8.660 & ~&9\\
\hline 
\end{tabular}
\caption{Exact excitation energies $E_{nm}$ of the two-electron model with hamiltonian specified in Eq.~(\ref{H2e}) in the non-interacting limit ($E_{nm}^0$, second column) and in the strongly correlated limit ($E_{nm}^\infty$, fourth column).    The eigenvalues of the QCM equations of motion ($E^\infty_{n+m-1,\pm}$)  in the strongly correlated regime are listed in the fifth column.  The top half of the table lists excitations with even displacement fields and the bottom half lists excitations with odd displacement fields. The third and the last column on the right list the number of nodes in the displacement field in the non interacting limit ($N^0$) and in the strongly correlated limit ($N^\infty$).}
\end{table}  
The elastic equation of motion can also be solved numerically, and the results are in very good agreement with the analytical solution.  This, and the fact that the sum rule~(\ref{SumRule}) is satisfied with good accuracy, builds our confidence in the numerical solution.

In Fig.~(\ref{EvolutionOfLevels}) we present the numerical results for some of the lowest-lying excitations as a function of $\omega_0$.  
We can immediately see that the ``non-degenerate" excitations, by which we mean the excitations $(1,0)$ $(0,1)$, and $(1,1)$, which are uniquely associated to a given displacement field,  are rather well reproduced by our calculation for all values of $\omega_0$ .  On the other hand, the ``degenerate excitations", which yield the same displacement field but have different energies, are replaced by a single excitation of an average energy, in such a way that the total spectral strength of the group is preserved.  Two examples of this phenomenon are evident in  Fig.~(\ref{EvolutionOfLevels}):   the $(2,0)$ and $(0,2)$ excitations, which  in the strong correlation limit ($\omega_0 \to 0$)  have energies $2 \omega_0$ and $3.464 \omega_0$ respectively, are replaced by a single excitation -- the fourth one in Fig.~(\ref{EvolutionOfLevels}) --  which  tends to the  ``average" energy $2.632  \omega_0$.  Similarly,  the $(3,0)$ and $(1,2)$ excitations, which, in the $\omega_0 \to 0$ limit tend to $3 \omega_0$ and $4.464 \omega_0$ respectively are replaced by a single excitation -- the fifth one in Fig.~(\ref{EvolutionOfLevels}) -- which tends to the ``average" energy $3.942 \omega_0$.    The pattern recurs for more complex multiplets of excitations, involving three or more states with the same displacement field and different energies.

\begin{figure}
\begin{center}
\includegraphics[width=1.0\linewidth]{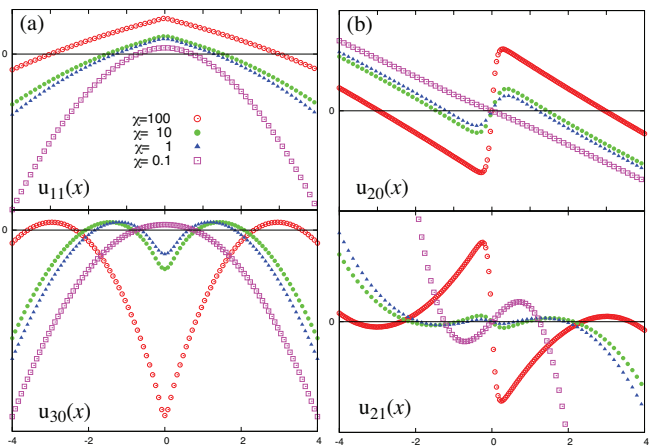}
\caption{(Color online)  Evolution of the displacement fields of excitations  $(1,1), (3,0)$ (even) and $(2,0),(2,1)$ (odd) as a function of correlation strength $\chi=\omega_0^{-1/2}$, as shown in the top left panel.  Notice the variation in the number of nodes as $\chi$ increases from the weakly correlated to the strongly correlated limit.} 
\label{EvolutionOfNodes}
\end{center}
\end{figure}

We notice that the displacement field  associated with, say, the $(n,m)$ excited state has a number of nodes that generally {\it grows} from $n+2m-1$ in the weak coupling limit to $2(n+m-1)$  (odd $n$) or $2(n+m-1)+1$ (even $n$)  in the strong coupling limit.   This effect is particularly pronounced for states of small $m$, and is absent in the $n=0$ states.   Fig.~(\ref{EvolutionOfNodes}) shows the evolution of the displacement field for the even excitations $(1,1)$  and  $(3,0)$   and for the odd excitations $(2,0)$  and  $(2,1)$.   We see that in the ``non-degenerate" $(1,1)$ state, the number of nodes stays constant  and equal to $2$ as one goes from the weakly correlated to the strongly correlated regime.  In the $(3,0)$ state the number of nodes grows from $2$ to $4$ nodes, so that the displacement field of this state becomes proportional, in the strong-correlation limit, to that of the much higher in energy $(0,3)$ state.   The same behavior is observed in state $(2,0)$, for which the number of nodes grows from $1$  to $3$, and in state $(2,1)$, for which it grows from  $3$ to $5$.  By this mechanism, states of very different energy end up sharing the same displacement field (up to a proportionality constant) in the strong correlation limit.

 Our discussion has been limited to singlet states (symmetric wave function in the relative channel).  It would be easy to extend the calculation to include triplet states.  To this end, we simply replace the density, kinetic energy density, and pair correlation function of the ground-state (a singlet)  by the same quantities calculated from the ground state in the triplet ($S=1$) sector of the Hilbert space.  The relative wave functions of these states are antisymmetric.   The correct symmetry of the wave function is automatically taken into account through the ground-state properties, and does not further appear in the elastic equation of motion.   

\section{Discussion and summary}
The elastic approximation is, in a very precise sense, the extension of the well-known collective approximation\cite{Collective,Pathak73} of the homogeneous electron gas to non-homogeneous electronic systems.
In the case of two electrons interacting by Coulomb potential in a harmonic trap, we have seen that the elastic approximation replaces groups of excitations characterized by the same displacement field by a single excitation that carries the oscillator strength of the whole group.   In more complex systems, we do not expect to be able to identify small groups of excitations that share the same displacement field.  All that can be said is that the displacements associated with different excitations will not be linearly independent.  Each eigenfunction of the elastic equation of motion will overlap with many different excitations. However, the integrated spectral strength of the elastic eigenmodes will still add up to the correct value.  For this reason, our approximation  should be useful in dealing with collective effects which depend on the integrated strength of the excitation spectrum, such as the dipolar fluctuations that are responsible for van-der Waals attraction.\cite{Langreth04,Dobson05}  Other possible applications include possible nonlocal refinements of the plasmon pole approximation in GW theory\cite{Onida02} and  studying the dynamics of strongly correlated systems, which are dominated by a collective response.  As a byproduct we got an explicit analytic representation of the exact xc kernel in the high-frequency (anti-adiabatic) limit.\cite{Nazarov10a} This kernel should help us to study an importance of the space and time nonlocalities in the KS formulation of TD(C)DFT. It would be particularly  interesting to try and interpolate between the adiabatic and anti-adiabatic extremes to construct a reasonable frequency-dependent functional.

The elastic equation of motion derived in this paper relied on the knowledge of the exact density matrices $\rho^{(1)}$ and $\rho^{(2)}$ of the ground-state.    In many cases, these ground-state properties can be extracted from Quantum Monte Carlo calculations.  When this cannot be done, one can still resort to density functional theory, i.e.  apply the QCM formulation directly to the Kohn-Sham system, in which case we 
do not need the exact ground-state density matrices, but 
only the  ground-state KS orbitals and a reasonable approximation for the exchange-correlation field.   While the standard KS method treats the noninteracting kinetic stress tensor exactly,  our method should be computationally more agile,  for large systems,  since it does not involve time-dependent orbitals and/or the inversion of large linear response matrices. 

It remains a challenge to extend the present formalism to the nonlinear regime, as well as including external magnetic fields and spin-orbit interactions.

\section{Acknowledgements}
This work was supported by DOE grant DE-FG02-05ER46203 (GV) and DE-AC52-06NA25396 (JT) and by the IKERBASQUE Foundation.  GX was supported by NSF of China under Grant No. 10704066 and 10974181. IVT acknowledges funding by the Spanish MEC (FIS2007-65702-C02-01), ``Grupos Consolidados UPV/EHU del Gobierno Vasco'' (IT-319-07), and the European Community through e-I3 ETSF project (Contract Number 211956). GV gratefully acknowledges the kind hospitality of the ETSF in San Sebastian where this work was completed.  We thank Dr. Stefano Pittalis for his help in calculating an plotting the curves shown in Fig.~\ref{OLTK}, and Dr. Paola Gori-Giorgi for kindly providing the code for calculating the structure factor of the electron gas.

\appendix

\section{Stress tensor operator}
From the evaluation of Eq.~(\ref{defP}) at the Euclidean metrics $g_{\mu\nu}=\delta_{\mu\nu}$ we get~\cite{tokatly2005a,tokatly2005b}
\be
\hat P_{\mu\nu}=\hat T_{\mu\nu}+\hat W_{\mu\nu}\,,
\ee
where
\be
\hat T_{\mu\nu}=\frac{1}{2m}\left\{(\partial_\mu \hat\Psi^\dagger)(\partial_\nu \hat\Psi)+(\partial_\nu \hat\Psi^\dagger)(\partial_\mu \hat\Psi)-\frac{1}{2}\nabla^2 \hat n \delta_{\mu\nu}\right\}\,,
\ee
and
\ber
\hat W_{\mu\nu}&=&-\frac{1}{2}\int d\rv' \frac{r'_\mu r'_\nu}{r'}\frac{\partial w(r')}{\partial r'}\nn\\
&\times&\int_0^1 d\lambda  \hat \rho_2(\rv+\lambda\rv',\rv-(1-\lambda)\rv')\,.
\eer
Here $\hat\Psi(\rv)$ is the field operator,  
\be
\hat \rho_2(\rv,\rv')= \hat\Psi^\dagger(\rv)\hat\Psi^\dagger(\rv')\hat\Psi(\rv') \hat\Psi(\rv) 
\ee
is the diagonal  two-particle density operator, and $w(r)$ is the interaction potential.

\section{Derivation of the force identity Eq.~(\ref{IlyasIdentity})}

In this appendix we derive an identity which is used in Sec.~II~A to identify the right hand sides of Eqs.~(\ref{F1}) and (\ref{IlyasIdentity}).  Namely, we consider a functional $S[g_{\mu\nu}]$ of the following metric tensor 
\be
\label{MetricTensor_App}
g_{\mu\nu}(\xiv,t)  = \frac{\partial r_{\alpha}}{\partial \xi_{\mu}}\frac{\partial r_{\alpha}}{\partial \xi_{\nu}}\,, \quad r_{\alpha}(\xiv)=\xi_{\alpha}+u_{\alpha}(\xiv)
\ee
and prove that the following equality holds
\begin{equation}
 \label{ForceIdentity}
\frac{\delta S}{\delta r_{\mu}} = - 2\frac{\partial}{\partial \xi_{\alpha}}\left(\frac{\partial r_{\mu}}{\partial \xi_{\beta}}\frac{\delta S}{\delta g_{\alpha\beta}}\right)\equiv
\frac{\partial}{\partial \xi_{\alpha}}\left(\frac{\partial r_{\mu}}{\partial \xi_{\beta}}\sqrt{g}P^{\alpha\beta}\right)
\end{equation}
Note that the identity relates the functional derivative of $S$ with respect to the displacement to the functional derivative with respect to the metric/deformation tensor, which physically means a connection of the force to the stress. 

To prove Eq.~(\ref{ForceIdentity}) we consider a small variation of the function $r_{\mu}(\xiv)$: $r_{\mu}(\xiv)\mapsto r_{\mu}(\xiv) + \delta r_{\mu}(\xiv)$. The corresponding variation of the functional $S$ takes the form
\begin{equation}
 \label{varS1}
\delta S = \int d\xiv \frac{\delta S}{\delta r_{\mu}}\delta r_{\mu}(\xiv)
\end{equation}
On other hand, the variation of  $r_{\mu}(\xiv)$ induces the following variation of the metric tensor: $g_{\mu\nu}(\xiv)\mapsto g_{\mu\nu}(\xiv) + \delta g_{\mu\nu}(\xiv)$, where
\begin{equation}
 \label{varg}
\delta g_{\mu\nu} = \frac{\partial \delta r_{\alpha}}{\partial \xi_{\mu}}\frac{\partial r_{\alpha}}{\partial \xi_{\nu}}
+ \frac{\partial r_{\alpha}}{\partial \xi_{\mu}}\frac{\partial \delta r_{\alpha}}{\partial \xi_{\nu}}
\end{equation}
Hence the variation of $S$ can be also written as
\begin{eqnarray}\nonumber
\delta S &=& \int d\xiv \frac{\delta S}{\delta g_{\mu\nu}}\delta g_{\mu\nu}(\xiv)\\
&=& \int d\xiv \frac{\delta S}{\delta g_{\mu\nu}} \left(\frac{\partial \delta r_{\alpha}}{\partial \xi_{\mu}}\frac{\partial r_{\alpha}}{\partial \xi_{\nu}}
+ \frac{\partial r_{\alpha}}{\partial \xi_{\mu}}\frac{\partial \delta r_{\alpha}}{\partial \xi_{\nu}}\right)
 \label{varS2a}
\end{eqnarray}
Performing the partial integration in the right hand side of Eq.~(\ref{varS2a}), and using the symmetry of the tensor $g_{\mu\nu}$ we reduce Eq.~(\ref{varS2a}) to the following form
\begin{equation}
 \label{varS2b}
\delta S = -\int d\xiv 2\frac{\partial}{\partial \xi_{\alpha}}\left(\frac{\partial r_{\mu}}{\partial \xi_{\beta}}\frac{\delta S}{\delta g_{\alpha\beta}}\right)\delta r_{\mu}(\xiv)
\end{equation}
The direct comparison of Eqs.~(\ref{varS1}) and (\ref{varS2b}) proves the announced identity ~(\ref{ForceIdentity}).

Finally, to make a connection to Eqs.~(\ref{F1}) and (\ref{IlyasIdentity}) we set $S[g_{\mu\nu}]=\hat{H}_u[g_{\mu\nu}]$ and take the expectation value of Eq.~(\ref{ForceIdentity}) in the state $|\tilde{\psi}(t)\rangle$. Then, linearizing with respect the displacement $\uv(\xiv)$, we find that the right hand side of Eq.~(\ref{ForceIdentity}) becomes identical to the right hand side of Eq.~(\ref{F1}), while the left hand side of Eq.~(\ref{ForceIdentity}) is exactly equal to the right hand side of Eq.~(\ref{IlyasIdentity}).

\section{Derivation of Eq.(\ref{T2uv})}

In this appendix we derive the linearized form of the kinetic energy Eqs.~(\ref{T2uv}) for the instantaneously distorted ground state.  

We start with the general nonlinear expression for the kinetic energy $T[{\bf u}]$ in the elastic approximation [see Eq.~(\ref{Tfull})]
\be
\label{Tfull_App}
T[\uv] = \frac{1}{2m}\int d\xiv  \sqrt{g} g^{\mu\nu}\partial_\mu\partial_\nu^\prime [g^{-1/4}(\rv) g^{-1/4}(\xiv') \rho(\xiv,\xiv')]_{\xiv=\xiv'}\,,
\ee
where $\rho(\xiv,\xiv')$ is the exact ground state one-particle density matrix, and $g^{\mu\nu}(\xiv)$ and $g(\xiv)$ are, respectively, the inverse and the determinant of the metric tensor $g_{\mu\nu}(\xiv)$ that is the functional of the displacement $\uv(\xiv)$, which is defined by Eq.~(\ref{MetricTensor_App})

Our aim is to expand the functional of Eq.~(\ref{Tfull_App}) to the second order in the displacement field, i.~e., to the first non-vanishing contribution corresponding to the linearized theory. 

First we explicitly calculate the derivatives in the right hand side of Eq.~(\ref{Tfull_App}) and set $\xiv'=\xiv$. As a result Eq.~(\ref{Tfull_App}) reduces to the form
\begin{eqnarray}
 \nonumber
&& T[\uv] = \int d\xiv g^{\mu\nu}\Big\{K_{\mu\nu} +\frac{n_0}{8m}(\partial_{\mu}\ln\sqrt{g})(\partial_{\nu}\ln\sqrt{g}) \\
&-& \frac{1}{8m}\big[(\partial_{\mu}\ln\sqrt{g})(\partial_{\nu}n_0)+(\partial_{\nu}\ln\sqrt{g})(\partial_{\mu}n_0)\big]  \Big\}\,,
\label{Tfull1}
\end{eqnarray}
where $n_0(\xiv)=\rho(\xiv,\xiv)$ is the ground state density, and
\begin{equation}
 \label{Kmunu}
K_{\mu\nu}(\xiv)=\frac{1}{2m}[\partial_{\mu}\partial_{\nu}'\rho(\xiv,\xiv')]_{\xiv=\xiv'}.
\end{equation}

Making use of the following representation for $\sqrt{g}$,
$$
\sqrt{g}=\det\left(\frac{\partial r_{\alpha}}{\partial\xi_{\beta}}\right)\,,
$$
we can write the derivative of $\ln\sqrt{g}$ as follows:
\begin{equation}
 \label{dlng}
\partial_{\mu}\ln\sqrt{g} = \partial_{\mu}\ln\det\left(\frac{\partial r_{\alpha}}{\partial\xi_{\beta}}\right) =
\frac{\partial \xi_{\alpha}}{\partial r_{\beta}}\partial_{\mu}\frac{\partial r_{\beta}}{\partial\xi_{\alpha}}\,.
\end{equation}
It is now straightforward to expand the right hand side of Eq.~(\ref{dlng}) to the second order in $\uv$:
\begin{equation}
 \label{dlng2}
\partial_{\mu}\ln\sqrt{g}\approx \partial_{\mu}\partial_{\alpha}u_{\alpha} - (\partial_{\beta}u_{\alpha})\partial_{\mu}\partial_{\alpha}u_{\beta}\,. 
\end{equation}
Next we consider the covariant tensor $g^{\mu\nu}$ (the inverse of $g_{\mu\nu}$):
\begin{equation}
 \label{g_cov}
g^{\mu\nu} = \big[\delta_{\mu\nu} + \partial_{\mu}u_{\nu} + \partial_{\nu}u_{\mu} + (\partial_{\mu}u_{\alpha})(\partial_{\nu}u_{\alpha})\big]^{-1}\,.
\end{equation}
Expanding the inverse matrix in Eq.~(\ref{g_cov}) to the second order and expressing the result in terms of the strain tensor $u_{\mu\nu}$ we get
\begin{equation}
 \label{g_cov2}
g^{\mu\nu}\approx \delta_{\mu\nu} -2u_{\mu\nu} + 4u_{\mu\alpha}u_{\nu\alpha} - (\partial_{\mu}u_{\alpha})(\partial_{\nu}u_{\alpha})\,.
\end{equation}
Finally, we substitute Eqs.~(\ref{dlng2}) and (\ref{g_cov2}) into Eq.~(\ref{Tfull1}) and keep terms up to the second order in $\uv$. The second order contribution to $T$ takes the form
\begin{widetext}
\begin{equation}
\label{T2_a}
T_2 = \int d\xiv \Big\{K_{\mu\nu}[4u_{\mu\alpha}u_{\nu\alpha} - (\partial_{\mu}u_{\alpha})(\partial_{\nu}u_{\alpha})] +
\frac{n_0}{8m}(\partial_{\mu}u_{\alpha\alpha})(\partial_{\mu}u_{\nu\nu}) 
+ \frac{\partial_{\nu}n_0}{2m}u_{\mu\nu}\partial_{\mu}u_{\alpha\alpha}  +  \frac{\partial_{\nu}n_0}{4m}(\partial_{\mu}u_{\alpha})\partial_{\nu}(\partial_{\alpha}u_{\mu})\Big\}\,.
\end{equation}
The last term in Eq.~(\ref{T2_a}) can be identically represented as follows
\begin{equation}
 \label{laplas}
\frac{\partial_{\nu}n_0}{4m}(\partial_{\mu}u_{\alpha})\partial_{\nu}(\partial_{\alpha}u_{\mu}) = 
-\frac{\nabla^2 n_0}{8m}[4u_{\mu\alpha}u_{\mu\alpha} - (\partial_{\mu}u_{\alpha})(\partial_{\mu}u_{\alpha})]
- \frac{\partial_{\nu}n_0}{2m}u_{\mu\alpha}\partial_{\nu}u_{\mu\alpha}\,. 
\end{equation}
Note that the coefficient in front of the first term in Eq.~(\ref{laplas}) and the corresponding coefficient in first term in Eq.~(\ref{T2_a}) are naturally combined into the kinetic stress tensor $T_{\mu\nu,0}=2K_{\mu\nu} - \delta_{\mu\nu}\nabla^2n_0/4m$. Hence, inserting Eq.~(\ref{laplas}) into Eq.~(\ref{T2_a}) and integrating by parts terms proportional to $\partial_{\nu}n_0/2m$ we arrive at the following final representation for the linearized kinetic energy of the distorted state:
\begin{equation}
\label{T2_b}
T_2 = \int d\xiv \Big\{\frac{1}{2}T_{\mu\nu,0}[4u_{\mu\alpha}u_{\nu\alpha} - (\partial_{\mu}u_{\alpha})(\partial_{\nu}u_{\alpha})] +
\frac{n_0}{8m}(\partial_{\mu}u_{\alpha\alpha})(\partial_{\mu}u_{\nu\nu}) 
+ \frac{n_0}{4m}[\partial_{\mu}u_{\nu\alpha}\partial_{\mu}u_{\nu\alpha}  - \partial_{\mu}u_{\nu\mu}\partial_{\mu}u_{\alpha\alpha}]\,, 
\end{equation}
\end{widetext}
which is identical to Eq.~(\ref{T2uv}). It is worth noting a convenient feature of this representation -- the last term in Eq.~(\ref{T2_b}) vanishes in all 1D systems and for homogeneous systems with $n_0=const$ in any number of dimensions.

\section{Proof of the orthonormality relation for single-particle transition currents}
In this appendix we prove Eq.~(\ref{1P-orthonormality}) for one-particle systems in the absence of a magnetic field.

The matrix elements of the current density operator is  
\be
\jv_{0n}(\rv)=\langle 0|\hat \jv(\rv)|n\rangle =-\frac{i}{2m} [\psi_0\nablabold\psi_n-\psi_n\nablabold \psi_0]\,,
\ee
where $\psi_0$,..., $\psi_n$ are orthonormal  eigenfunctions of the one-electron hamiltonian, which can be assumed to be real if there is no magnetic field.   Now observe that
\be
\frac{\jv_{0n}(\rv)}{\sqrt{n_0(\rv)}} = -\frac{i}{2m} \frac{\psi_0\nablabold\psi_n-\psi_n\nablabold\psi_0}{\psi_0}=-\frac{i}{2m}\psi_0 \nablabold \left(\frac{\psi_n}{\psi_0}\right)\,,
\ee
and $\jv_{n0}(\rv)=-\jv_{0n}(\rv)$.
Also, from the continuity equation we get
\be
\nablabold\cdot\left(\psi_0\nablabold\psi_n-\psi_n\nablabold\psi_0\right)=-2m\omega_{n0}\psi_0\psi_n\,.
\ee
Combining these two equations we get
\be
\nablabold\cdot\left(\psi_0^2\nablabold\left(\frac{\psi_n}{\psi_0}\right)\right)=-2m\omega_{n0}\psi_0\psi_n\,.
\ee
From this we see that
\ber
&&\int d\rv \left(\sqrt {\frac{2m}{\omega_{n0}}} \frac{\jv_{0n}(\rv)}{\sqrt{n_0(\rv)}} \right) \cdot \left(\sqrt{\frac{2m}{\omega_{k0}}}\frac{\jv_{k0}(\rv)}{\sqrt{n_0(\rv)}}\right) =
\nn\\
&=&\frac{1}{2m}\frac{1}{\sqrt{\omega_{n0}\omega_{k0}}} \int d\rv  \left[\nablabold\left(\frac{\psi_n}{\psi_0}\right)\right] \cdot \psi_0^2(\rv) \nablabold\left(\frac{\psi_k}{\psi_0}\right) \nn\\
&=&-\frac{1}{2m}\frac{1}{\sqrt{\omega_{n0}\omega_{k0}}}\int d\rv  \frac{\psi_n}{\psi_0}\nablabold\cdot\left[\psi_0^2(\rv)\nablabold\left(\frac{\psi_k}{\psi_0}\right)\right]\nn\\
&=&\sqrt{\frac{\omega_{k0}}{\omega_{n0}}}\int d\rv \psi_n \psi_k \nn\\
&=&\delta_{nk}\,
\eer
which proves Eq.~(\ref{1P-orthonormality}).

\section{Proof of the sum rules (\ref{SumRule}) - (\ref{OscillatorStrengthSumRule})}
 Our starting point is the first moment sum rule for the current-current response function (or ``third-moment sum rule" for the density-density response function), which states that
\be\label{FirstMoment}
-\frac{1}{\pi}\int_0^\infty d \omega \omega \Im m\chi_{\mu\nu}(\rv,\rv',\omega)=\sum_n \omega_{n0} [j_\mu(\rv)]_{0n} [j_\nu(\rv')]_{n0}\,.
\ee

On the other hand, the equation of motion~(\ref{eom.last}) for the current density in  the elastic approximation can be rewritten as
\ber
&&\int d\rv' \left\{\omega^2 \delta_{\mu\nu}\delta(\rv-\rv')-\tilde M_{\mu\nu}(\rv,\rv')\right\} \sqrt{\frac{m}{n_0(\rv')}}j_\nu(\rv') \nn\\
&&= \omega^2 \sqrt{\frac{n_0(\rv)}{m}}A_{1,\mu}(\rv)\,,
\eer
where $A_{1,\mu}=\partial_{\mu}V_1/(i\omega)$, $j_\nu=-i\omega n_0 u_\nu$, and
\ber\label{EquationE3}
\tilde M_{\mu\nu}(\rv,\rv') &=& 
\frac{1}{\sqrt{m n_0(\rv)}}M_{\mu\nu}(\rv,\rv')\frac{1}{\sqrt{m n_0(\rv')}}\nn\\
&=&\sqrt{\frac{m}{n_0(\rv)}}\sum_n\omega_{n0} \left\{[j_\mu(\rv)]_{0n} [j_\nu(\rv')]_{l0}\right.\nn\\
&+&\left.[j_\nu(\rv')]_{0n} [j_\mu(\rv)]_{n0}  \right\}\sqrt{\frac{m}{n_0(\rv')}}\,.
\eer
is a hermitian positive definite operator, which admits a complete set of orthonormal eigenfunctions.  Let us denote by $\tilde \uv_\lambda(\rv)$ these eigenfunctions, and by $\omega_\lambda^2$ their eigenvalues.
The orthonormality relation reads
\be
\int d\rv \tilde \uv_\lambda(\rv) \cdot \tilde \uv_{\lambda'}(\rv)  = \delta_{\lambda\lambda'}
\ee
and the completeness relation is
\be\label{EquationE5}
\sum_\lambda [\tilde \uv_\lambda(\rv)]_\mu [\tilde \uv_\lambda(\rv')]_\nu=\delta(\rv-\rv')\delta_{\mu\nu}\,.
\ee
The kernel itself can be written as
\be
\tilde M_{\mu\nu}(\rv,\rv')  = \sum_\lambda \omega_\lambda^2 [\tilde \uv_\lambda(\rv)]_\mu [\tilde \uv_\lambda(\rv')]_\nu\,. 
\ee

 The equation of motion for the current density can be rewritten as
\ber
&&\sum_\lambda \int d\rv' (\omega^2-\omega_\lambda^2) [\tilde \uv_\lambda(\rv)]_\mu [\tilde \uv_\lambda(\rv')]_\nu \sqrt{\frac{m}{n_0(\rv')}}j_\nu(\rv') \nn\\
&=& \omega^2 \sqrt{\frac{n_0(\rv)}{m}}A_\mu(\rv)\,,
\eer
Its solution is obtained by projecting both sides of the equation along the eigenvector $\tilde \uv_\lambda$.  We get
\be
\left[\sqrt{\frac{m}{n_0}}\jv\right]_\lambda =  \left(1 +\frac{\omega_\lambda^2}{\omega^2-\omega_\lambda^2}\right)\left[\sqrt{\frac{n_0}{m}}\Av\right]_\lambda\,,
\ee
where the subscript $\lambda$ denotes projection along $\tilde \uv_\lambda$.  From this we obtain
\ber
j_\mu(\rv)&=& \frac{n_0(\rv)}{m}A_\mu(\rv) +\int d\rv'\sum_\lambda\frac{\omega_\lambda^2}{\omega^2-\omega_\lambda^2}\sqrt{\frac{n_0(\rv)}{m}} [\tilde \uv_\lambda(\rv)]_\mu \nn\\
&&[\tilde \uv^\lambda(\rv')]_\nu\sqrt{\frac{n_0(\rv')}{m}}A_\nu(\rv')\,.
\eer
Hence, the current-current response function in the elastic approximation is
\ber
&&\chi^{el}_{\mu\nu}(\rv,\rv',\omega)=\frac{n_0(\rv)}{m}\delta_{\mu\nu}\delta(\rv-\rv')\nn\\
&&+\sum_\lambda\frac{\omega_\lambda^2}{\omega^2-\omega_\lambda^2}\sqrt{\frac{n_0(\rv)}{m}} [\tilde \uv_\lambda(\rv)]_\mu  [\tilde \uv_\lambda(\rv')]_\nu\sqrt{\frac{n_0(\rv')}{m}}\,.\nn\\
\eer
Evaluating the sum rule we obtain
\ber
&&-\frac{1}{\pi}\int_0^\infty d \omega \omega \Im m\chi^{el}_{\mu\nu}(\rv,\rv',\omega)=\nn\\
&&\frac{1}{2}\sum_\lambda \omega_\lambda^2 \sqrt{\frac{n_0(\rv)}{m}}[\tilde \uv_\lambda(\rv)]_\mu [\tilde \uv_\lambda(\rv')]_\nu\sqrt{\frac{n_0(\rv')}{m}}=\nn\\
&&\frac{1}{2m} \sqrt{n_0(\rv)}\tilde M_{\mu\nu}(\rv,\rv')\sqrt{n_0(\rv')}
\eer

On the other hand, the first moment sum rule~(\ref{FirstMoment})  can be rewritten as
\ber
&&-\frac{1}{\pi}\int_0^\infty d \omega \omega \Im m\chi_{\mu\nu}(\rv,\rv',\omega)=\nn\\
&&\frac{1}{2m}\sqrt{n_0(\rv)}\tilde M_{\mu\nu}(\rv,\rv')\sqrt{n_0(\rv')}\,.
\eer
Comparing the last two equations we conclude that 
\ber
&&-\frac{1}{\pi}\int_0^\infty d \omega \omega \Im m\chi_{ik}(\rv,\rv',\omega)=\nn\\&&-\frac{1}{\pi}\int_0^\infty d \omega \omega \Im m\chi^{el}_{ik}(\rv,\rv',\omega)\,,
\eer
i.e. the sum rule is satisfied in the elastic approximation.

More pointedly, making use of Eq.~(\ref{EquationE3})  the sum rule can be written in the form
\be
\sum_\lambda \omega_\lambda^2 [\tilde \uv_\lambda(\rv)]_\mu [\tilde \uv_\lambda(\rv')]_\nu=2m\sum_n \omega_{n0} \frac{[j_\mu(\rv)]_{0n}}{\sqrt{n_0(\rv)}} \frac{ [j_\nu(\rv')]_{n0}}{\sqrt{n_0(\rv')}}\,,
\ee
from which it follows that
\be
\omega_\lambda^2 = \sum_n \omega_{n0}^2 f^\lambda_n\,,
\ee
where the ``oscillator strengths" $f^\lambda_n$ are positive quantities defined as
\be
f^\lambda_n=\frac{2m |F^\lambda_n|^2}{\omega_{n0}}
\ee
with
\be
F^\lambda_n \equiv \int d\rv \tilde \uv_\lambda(\rv) \cdot \frac{[\jv(\rv)]_{0n}}{\sqrt{n_0(\rv)}}
\ee

As a final step we prove that
\be
\sum_l f^\lambda_n = 1
\ee
for all $\lambda$.
This is of course nothing but the $f$-sum rule
\be
-\frac{1}{\pi}\int_0^\infty d \omega \frac{\Im m\chi_{\mu\nu}(\rv,\rv',\omega)}{\omega}= \frac{n_0(\rv)}{2m}\delta_{\mu\nu}\delta(\rv-\rv')\,,
\ee
which is manifestly satisfied by $\chi_{\mu\nu}^{el}$ by virtue of the completeness relation~(\ref{EquationE5}) for $\tilde \uv_\lambda(\rv)$.  
When applied to the exact response function the $f$-sum rule implies that
\ber
\sum_n\frac{[j_\mu(\rv)]_{0n}[j_\nu(\rv')]_{n0}}{\omega_{n0}} =\frac{n_0(\rv)}{2m}\delta_{\mu\nu}\delta(\rv-\rv')\,.
\eer
Then we see that
\ber
&&\sum_n\frac{2m |F^\lambda_n|^2}{\omega_{n0}}= 2m \int d\rv \int d\rv'\nn\\
&&  \sum_n \frac{[j_\mu(\rv)]_{0n}[j_\nu(\rv')]_{n0}}{\omega_n}\frac{[\tilde \uv_\lambda(\rv)]_\mu}{\sqrt{n_0(\rv)}} \frac{[\tilde \uv_\lambda(\rv')]_\nu}{\sqrt{n_0(\rv')}}\nn\\
&&=\int d\rv \int d\rv' n_0(\rv) \delta_{\mu\nu}\delta(\rv-\rv')\frac{[\tilde \uv_\lambda(\rv)]_\mu}{\sqrt{n_0(\rv)}}\frac{[\tilde \uv_\lambda(\rv')]_\nu}{\sqrt{n_0(\rv')}}\nn\\
&& =1\,.
\eer
QED.

\section{Analytic solution of the elastic eigenvalue problem in the strong-correlation regime}

In this appendix we present an asymptotically exact solution of the 1D continuum mechanics eigenvalue problem for two particles confined by the harmonic potential $V_0=\frac{1}{2}m\omega_0^2x^2$ and interacting with a soft-Coulomb potential
\begin{equation}
\label{softC}
w(x-x')=\frac{e^2}{\sqrt{(x-x')^2+a^2}}\,.
\end{equation}
At the exact many-body theory level the system is described by the Hamiltonian of Eq.~(\ref{H2e}). Within our continuum mechanics the excitation energies are obtained from the solution of the following  ``elastic" eigenvalue problem for the displacement $u(x,t)=u(x)e^{-i\omega t}$ (see Eq.~(\ref{eom.1dimension}) in the main text)
\begin{widetext}
\begin{equation}
 \label{Eq1}
m\omega^2n_0 u(x) = \frac{1}{4m}\partial_x^2~[n_0\partial_x^2u(x)] - 3\partial_x[T_0\partial_xu(x)]
+ m\omega_0^2n_0u(x) + 2\int dx'[\partial_x^2w(x-x')]\Psi_{0}^2(x,x')[u(x)-u(x')]\,.
\end{equation}
Here $\Psi_{0}(x,x')$ is the ground state two-particle wave function, which in this case coincides with the square root of the two-particle density matrix, $n_0(x)$ is the ground state density, and $T_0(x)$ is the ground state kinetic stress tensor defined by Eq.~(\ref{T01D}). 

Equation (\ref{Eq1}) possesses an analytic solution in the limit of strong Coulomb interaction $e^2m^2\omega_0\gg 1$ when the ground state wave function reduces to the following asymptotic form
\begin{equation}
 \label{Psi0}
\Psi_0(x_1,x_2)=\frac{1}{\ell_{\infty}\sqrt{\pi}}e^{-\frac{(x_1+x_2)^2}{2\sqrt{3}\ell_{\infty}^2}}
\left[e^{-\frac{(x_1-x_2-x_0)^2}{2\ell_{\infty}^2}}+e^{-\frac{(x_1-x_2+x_0)^2}{2\ell_{\infty}^2}}\right]\,,
\end{equation}
\end{widetext}
where $\ell_{\infty}=(2\hbar/\sqrt{3}m\omega_0)^{\frac{1}{2}}$, and $x_0=(2e^2/m\omega_0^2)^{\frac{1}{3}}$ is the classical distance between particles, i.e., the distance that minimizes the classical energy of two charged particles in the 1D harmonic potential.
The corresponding ground state density takes the form of two well separated ``blobs"
\begin{equation}
 \label{n0}
n_0(x) = \frac{1}{\lambda_{\infty}\sqrt{\pi}}\left[e^{-\frac{(x-x_0/2)^2}{\lambda_{\infty}^2}}
+ e^{-\frac{(x+x_0/2)^2}{\lambda_{\infty}^2}}\right]\,,
\end{equation}
where the size $\lambda_{\infty}$ of the density blobs located at $x=\pm x_0/2$ is
\begin{equation}
 \label{lambda}
\lambda_{\infty} = \left(\frac{\hbar(\sqrt{3}+1)}{2\sqrt{3}m\omega_0}\right)^{\frac{1}{2}}\,.
\end{equation}
The kinetic stress tensor Eq.~(\ref{T01D}) for the ground state wave function (\ref{Psi0}) becomes simply proportional to the density:
\begin{equation}
 \label{T0}
T_0(x) = \frac{\sqrt{3}+1}{4}\omega_0 n_0(x)\,.
\end{equation}

Another technical observation which simplifies calculations in the strong interaction limit is that the cross-products of the two exponentials in the square brackets in (\ref{Psi0}) are irrelevant for the expressions of the type $\Psi_{0}(x,x_2)\Psi_{0}(x',x_2)$ in the limit of $x_0\sqrt{m\omega_0}\gg 1$. For the two-particle density matrix entering the nonlocal term in (\ref{Eq1}) this implies the following result
\begin{widetext}
\begin{equation}
 \label{rho2}
2\Psi_{0}^2(x,x') =\frac{2}{\ell_{\infty}^2\pi}e^{-\frac{(x+x')^2}{\sqrt{3}\ell_{\infty}^2}}
\left[e^{-\frac{(x-x'-x_0)^2}{\ell_{\infty}^2}}+e^{-\frac{(x-x'+x_0)^2}{\ell_{\infty}^2}}\right]\,.
\end{equation}
\end{widetext}

Simplification of the integral term in the equation of motion (\ref{Eq1}) comes from the fact that in the limit of $x_0\sqrt{m\omega_0}\gg 1$ the pair correlation function (\ref{rho2}) is peaked at $|x-x'|\sim x_0$ (this keeps the particles at a distance close to the classical value). Hence in the integral kernel the interaction factor can be approximated as
\begin{equation}
 \label{interaction}
\partial_x^2w(x-x')\approx \frac{2e^2}{|x-x'|^3} \approx \frac{2e^2}{x_0^3} = m\omega_0^2\,.
\end{equation}
Substituting (\ref{T0}) and (\ref{interaction}) into the (\ref{Eq1}) and using the obvious identity
$$
n_0(x) = \int dx'2\Psi_{0}^2(x,x')\,,
$$
we reduce the equation of motion to the following form
\begin{widetext}
\begin{equation}
m\left[\omega^2 - 2\omega_0^2\right]n_0(x) u(x) =  -m\omega_0^2\int dx'2\Psi_{0}^2(x,x')u(x')
+\frac{1}{4m}\partial_x^2 [n_0(x)\partial_x^2u(x)]- 
3\omega_0\frac{\sqrt{3}+1}{4}\partial_x[n_0(x)\partial_xu(x)]\,.
\label{Eq2}
\end{equation}
\end{widetext}

Now the following observations are in order:

(i) The integro-differential operator in (\ref{Eq2}) (in fact in the original Eq.~(\ref{Eq1})) is  symmetric under inversion of $x$. Hence the solutions can be classified by parity $u^{\pm}(x)=\pm u^{\pm}(-x)$. Therefore it is sufficient to consider Eq.~(\ref{Eq2}) only in the region of positive $x$;

(ii) In (\ref{Eq2}) for $x>0$ all local terms contain $n_0(x)$ which is a narrow Gaussian located at $x\sim x_0/2$. Therefore these terms are nonzero only around $x_0/2$;

(iii) The integral kernel $\Psi_{0}^2(x,x')$ for $x>0$ is a product of two Gaussian peaks, one at $x+x'\sim 0$, and another at $x-x'\sim x_0$, which confines $x$ to the region of the right density blob, $x\sim x_0/2$, and $x'$ to the region of the left blob, $x'\sim -x_0/2$.

Therefore for positive $x$ all terms in (\ref{Eq2}) are nonzero only in the region $x\sim x_0/2$, while the integration region in the nonlocal (interaction) term is confined by the Gaussian factors to $x'\sim -x_0/2$. Note that for this reason the integral term will contribute with opposite signs to the equations of motion for the modes of opposite parity.

To further simplify the eigenvalue problem in the strong coupling limit we proceed as follows.
(i) Considering Eq.~(\ref{Eq2}) in the region $x>0$ we make a shift of coordinates $x\to x + x_0/2$, and $x'\to x' -x_0/2$. After that because of the Gaussian factors the integration can be extended to the whole axis. This completely eliminates $x_0$ (i.e. the coupling constant) from the problem, as it should be in the strong coupling limit;
(ii) Go to dimensionless coordinates $\xi=x/\lambda_{\infty}$, and $\xi'= x'/\lambda_{\infty}$;
(iii) Divide everything by the ground state density $n_0$ (which is simply a Gaussian located at the origin after the above shift, $n_0(\xi)=e^{-\xi^2}/\sqrt{\pi}$).

The result of these three steps is the following dimensionless equation of motion
\begin{widetext}
\begin{equation}
\left(\frac{\omega^2}{\omega_0^2} -2\right)u^{\pm} = \frac{3e^{\xi^2}\partial^2[e^{-\xi^2}\partial^2 u^{\pm}]}{(\sqrt{3}+1)^2}
-\frac{3\sqrt{3}}{2}e^{\xi^2}\partial[e^{-\xi^2}\partial u^{\pm}]
\mp \frac{\sqrt{3}+1}{2\sqrt{\pi}3^{1/4}}\int\limits_{-\infty}^{\infty}d\xi'
e^{-\frac{(\sqrt{3}+1)^2}{4\sqrt{3}}\left(\xi' + \frac{\sqrt{3}-1}{\sqrt{3}+1}\xi \right)^2}u^{\pm}(\xi')\,.
\label{Eq3}
\end{equation}

To solve the eigenvalue problem (\ref{Eq3}) we employ the following identities for Hermite polynomials\cite{Gradshteyn}
\begin{eqnarray}
 \label{HermiteA}
&& e^{x^2}\partial_x\left(e^{-x^2}\partial_x H_k\right) = -2kH_k\,, \quad
e^{x^2}\partial_x^2\left(e^{-x^2}\partial_x^2 H_k\right) = 4k(k-1)H_k\,,\\
\label{HermiteB}
&& \frac{a+1}{2\sqrt{a\pi}}\int\limits_{-\infty}^{\infty}dx'
e^{-\frac{(a+1)^2}{4a}\left(x' + \frac{a-1}{a+1}x \right)^2}H_k(x') =(-1)^k\left(\frac{a-1}{a+1}\right)^kH_k(x)\,.
\end{eqnarray}

From the identities (\ref{HermiteA})and (\ref{HermiteB}) we see that Hermite polymomials are the eigenfunctions of each of three terms in the integro-differential operator (both for odd and for even modes) on the right hand side in (\ref{Eq3}). The corresponding eigenvalues (one should apply the identity (\ref{HermiteB}) with $a=\sqrt{3}$) are 
\begin{equation}
 \label{Omega}
\left(\frac{\omega_k^{\pm}}{\omega_0}\right)^2 = 2 + 3\frac{4k(k-1)}{(\sqrt{3}+1)^2}+ 3\sqrt{3}k \mp 
(-1)^k\left(\frac{\sqrt{3}-1}{\sqrt{3}+1} \right)^k\,,
\end{equation}
where $k=0,1,2,\dots$ is the quantum number labeling the eigenmodes (for every $n$ there are two modes of opposite parity). Note that the second, third, and fourth terms on the right hand side of Eq.~(\ref{Omega}) are, respectively, the eigenvalues of the first, second, and the third terms on the right hand side in Eq.~(\ref{Eq3}). With a little algebra we simplify the eigenvalues of Eq.~(\ref{Omega}) as follows
\be
\label{Omega_fin}
\omega_k^{\pm} = \omega_0\left[2+3\sqrt{3}k +6k(k-1)(2-\sqrt{3}) \mp (-1)^k(2-\sqrt{3})^k\right]^{1/2}\,.
\ee

In the physical units of length the eigenfunctions in the whole space take the form
\begin{equation}
 \label{modes}
n_0(x)u_k^{\pm}(x) \sim e^{-\frac{(x-x_0/2)^2}{\lambda_{\infty}^2}}H_k\left(\frac{x-x_0/2}{\lambda_{\infty}}\right) \pm
(-1)^{k}e^{-\frac{(x+x_0/2)^2}{\lambda_{\infty}^2}}H_k\left(\frac{x+x_0/2}{\lambda_{\infty}}\right)\,.
\end{equation}
Finally, the displacement eigenmodes normalized by the condition
\begin{equation}
 \label{normal}
\int dx n_0(x)u_k^{\pm}(x)u_l^{\pm}(x) = \delta_{kl}
\end{equation}
can be written as follows
\begin{equation}
 \label{Nmodes}
u_k^{\pm}(x) = \frac{1}{\sqrt{2^{k+1}k!}}\frac{e^{-\frac{(x-x_0/2)^2}{\lambda_{\infty}^2}}H_k\left(\frac{x-x_0/2}{\lambda_{\infty}}\right) \pm (-1)^{k}
e^{-\frac{(x+x_0/2)^2}{\lambda_{\infty}^2}}H_k\left(\frac{x+x_0/2}{\lambda_{\infty}}\right)}{e^{-\frac{(x-x_0/2)^2}{\lambda_{\infty}^2}}+e^{-\frac{(x+x_0/2)^2}{\lambda_{\infty}^2}}}\,.
\end{equation}
\end{widetext}
Equations (\ref{Omega_fin}) and (\ref{Nmodes}) give the asymptotically exact solutions of the elastic eigenvalue problem in the limit of strong correlations. In Sec.~IV~C we have used this solutions to control the accuracy of our numerical results.

\end{document}